\documentclass[english,11pt,reqno]{smfart}

\usepackage{amssymb}
\usepackage{amsmath}
\usepackage{amsthm}
\usepackage{dsfont}
\usepackage{graphicx}
\usepackage{subcaption}
\usepackage{appendix}
\usepackage{graphicx}
\usepackage{a4wide}
\usepackage{wasysym}
\usepackage{textcomp}
\usepackage{pgf,tikz}
\usetikzlibrary{arrows}
\usepackage{bm} 
\usepackage{gnuplottex}
\usepackage{gnuplot-lua-tikz}
\usepackage[french]{babel}
\usepackage{pgfplots, tikz}

\usetikzlibrary{mindmap,trees}
\usetikzlibrary{positioning}
\usetikzlibrary{decorations.text,decorations.pathmorphing}
\usetikzlibrary{shapes}
\usetikzlibrary{shadows}
\usetikzlibrary{decorations.markings}
\usetikzlibrary{calc}

\definecolor{myblue}{HTML}{92dcec}
\tikzstyle{every annotation}=[fill=myblue, font=\sf]

\newtheorem{theorem}{Theorem}
\newtheorem{proposition}[theorem]{Proposition}
\newtheorem{lemma}[theorem]{Lemma}

\newtheorem{definition}[theorem]{Definition}
\newtheorem{remark}{Remark}

\def\di{\displaystyle}

\newcommand{\N}{\mathbb{N}}

\newcommand{\R}{\mathbb{R}}

\newcommand{\T}{\mathbb{T}}
\newcommand{\TT}{\mathbb{T}}
\newcommand{\PTT}{\pmb \TT}

\newcommand{\fonctionsansdef}[3]{\begin{array}[t]{lrcl}#1 :&#2 &\longrightarrow &#3 \end{array}}

\def\TK{\mathbb{T}^\kappa}
\def\Tk{\mathbb{T}_\kappa}
\def\TKk{\mathbb{T}^\kappa_\kappa}

\def\Crd{C^0_{\mathrm{rd}}}

\definecolor{ffqqtt}{rgb}{1.,0.,0.2}
\definecolor{qqccqq}{rgb}{0.,0.8,0.}
\definecolor{ffqqqq}{rgb}{1.,0.,0.}
\definecolor{qqqqff}{rgb}{0.,0.,1.}

\begin{document}
\setcounter{tocdepth}{3}
\title[Multiscale functions, Scale dynamics and applications]{Multiscale functions, Scale dynamics and applications to partial differential equations}
\author{Jacky Cresson \and Fr\'ed\'eric Pierret}

\maketitle

\begin{abstract} 
Modeling phenomena from experimental data, always begin with a \emph{choice of hypothesis} on the observed dynamics such as \emph{determinism}, \emph{randomness}, \emph{derivability} etc. Depending on these choices, different behaviors can be observed. The natural question associated to the modeling problem is the following : \emph{``With a finite set of data concerning a phenomenon, can we recover its underlying nature ?} From this problem, we introduce in this paper the definition of \emph{multi-scale functions}, \emph{scale calculus} and \emph{scale dynamics} based on the \emph{time-scale calculus} (see \cite{bohn}). These definitions will be illustrated on the \emph{multi-scale Okamoto's functions}. The introduced formalism explains why there exists different continuous models associated to an equation with different \emph{scale regimes} whereas the equation is \emph{scale invariant}. A typical example of such an equation, is the \emph{Euler-Lagrange equation} and particularly the \emph{Newton's equation} which will be discussed. Notably, we obtain a \emph{non-linear diffusion equation} via the \emph{scale Newton's equation} and also the \emph{non-linear Schr\"odinger equation} via the \emph{scale Newton's equation}. Under special assumptions, we recover the classical \emph{diffusion} equation and the \emph{Schr\"odinger equation}.
\end{abstract}

\tableofcontents

\section{Introduction}

This article deals with new mathematical tools to deal with scale phenomena and applications to partial differential equations. The framework that we have developed can be read from a mathematical point of view following each definitions and theorems. However, this framework can be seen as a synthesis of different tentative of one of the authors in order to deal with scales in geometry and analysis in the context of different physical problems (see \cite{cresson2003,cresson2006,cresson2005,cresson2012,cresson_greff}), in particular the {\it scale relativity} theory developed by L. Nottale \cite{nottale1,nottale2,nottale3}, and more generally modelling problems. As a consequence, before coming to more mathematical considerations, we picture some important problems in modelling in Physics which are underlying our framework.\\ 

Modelling a given phenomenon from experimental data using classical mathematical tools always assume, sometimes implicitly, a given \emph{framework hypothesis} on the {\it real nature} of the phenomenon which can be also called the {\it texture of reality}. As an example, classical mechanics is developed using the classical differential calculus to write speed and acceleration of particle and implicitly assuming that the behaviour of these particles can be described using smooth curves on a given space. Depending on this framework, different behaviors will be predicted or not and will be confronted to reality. However, this assumption about the real nature of a phenomenon is in general not so easy to decide and in some sense depends on philosophical considerations (positivism, etc) which can not be proved. The classical debate between A. Einstein and N. Bohr about the nature of quantum physics is a famous example.\\

The previous problem can be handled using a different approach, looking at the way mathematical models for a given phenomenon are constructed. Indeed, the framework question is in fact related to two different facts which are in general mixed in the literature. In order to put in evidence these points, we first remind very roughly the usual way to construct a model for a given phenomenon :
\begin{itemize}
	\item Acquiring experimental data.
	\item Computations of relevant quantities (velocity, acceleration, etc).
	\item Functional relation between these quantities (at a discrete level).
	\item Asymptotic passage to a continuous model under a specific choice of hypothesis.
 \item Comparison to reality using numerical simulations
\end{itemize}

Putting apart the last step for the moment, we see that the framework assumption has to do with the following points :

\begin{itemize}
\item {\it Scale dependence} : Experimental data are intrinsically scale dependent via the measurement apparatus which induces scale of observation. This remark is well known but the mathematical framework to deal with it is a priori not developed. We return to a more pragmatic way to deal with this question by introducing the notion of \emph{scale functions} which are basically infinite family of discrete functions define on a given time-scale. All these objects and notions are based on the \emph{time-scale calculus} introduced in the late 88 by S. Hilger (see \cite{bohn}) in order to unify the classical continuous analysis and discrete calculus of finite differences. 

\item {\it Asymptotic behaviour} : The second remark is that the framework assumption concerns the asymptotic behavior (if any) of a scale dependent function. This is clearly something which is not discussed in the literature because the asymptotic procedure used is not put in evidence and implicitly assumed. The usual way is precisely to begin with an asymptotic object which is designed in a given framework and to check and rely on observed quantities with a given scale corresponding to the precision of measurement. In this paper, we define scale equations and an asymptotic procedure on scale functions based on the notion of scale regime which allows us to define a natural asymptotic object to a scale dependent one, a scale regime being more or less a stable scale behavior of a given scale function. 
\end{itemize}

As a consequence, the framework assumption or texture of reality is a fluctuating notion. It depends on the observed scale regime and the initial scale equation. As an example, classical mechanics corresponds to a linear scale regime and a part of quantum mechanics to a fractional one. We then recover different continuous objects representing a given phenomenon on different scale regime. 

It must be pointed out that this result is not trivial. In general, having a given framework, a continuous model is written and must, by construction, cover all the scale of observation. Doing so, for a given scale, one must understand how some quantities can be negligible. 

In our setting, a given scale regime will modify the corresponding continuous models, making some perturbations terms to appear or disappear. A new understanding of the way some equations can be seen as bifurcation or perturbation of some other is then possible. \\ 

This approach, although natural with respect to the modeling problem, leads to the following question : {\bf If all the quantities and equations are scale dependent does it means that no universal principle can be derived to described the asymptotic equations describing a particular phenomenon ?}  As Physics is based on the search for universal laws or principle, this problem is fundamental. The answer is fortunately no. A universal principle is a principle which can organize all scales from a given one. An example of such a tentative is given by the {\it scale relativity principle} initiated by L. Nottale \cite{nottale1,nottale2,nottale3} which was, as we already mentioned, the inspiration for part of this work. This principle states that the equations of motion correspond to some critical point of a Lagrangian functional at all scales or in other words that they keep an Euler-Lagrange form. This sentence can be rigorously defined using the notion of scale invariant equations in the formalism of the scale calculus that we introduce following our previous works \cite{cresson,cresson2003,cresson2005,cresson_greff,cresson2012} which is based on {\it embedding formalism} initiated in \cite{cresson,cd}. We are in particular able to define natural analogue of classical notions such as Lagrangian, functional, Hamiltonian, symmetries, first integrals, etc.  

\begin{figure}[!h]
	\centering
	\resizebox{0.5\textwidth}{!}{\includegraphics{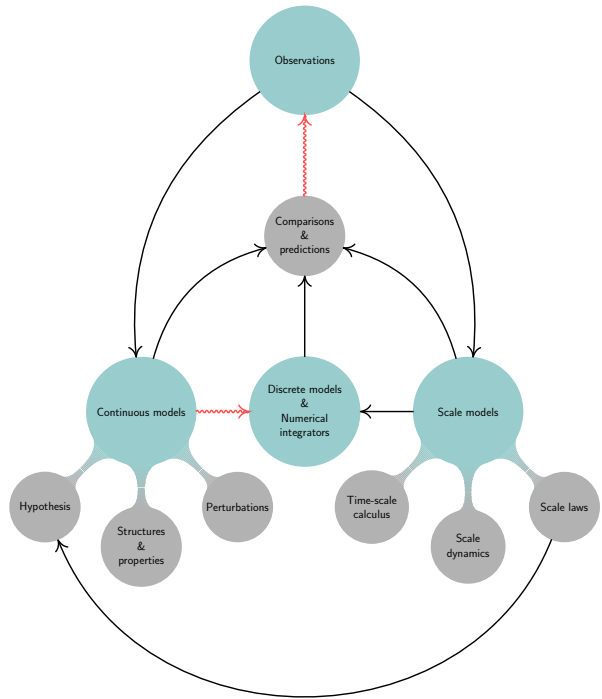}}
	\caption{Modeling problem. The red connection means that this is dependent of explicit choices.}
	\label{figure_resume}
\end{figure}

In this paper, we define what we call multiscale functions which are the geometrical objects underlying all modeling problems and the analysis tools necessary to study the behavior of these objects under change of scales. These definitions are connected with our previous attempt to define {\it scale manifold} in \cite{cresson2006} for the geometric part and the scale calculus introduced in \cite{cresson2003,cresson2005} and further developed in \cite{cresson_greff} for the analysis part. 

Our constructions and definitions use the framework of \emph{time-scale calculus} initiated by S. Hilger \cite{hilger88,hilger} in 1988 and further developed in \cite{bohn}. The time-scale calculus unifies the classical and discrete calculus in the sense that one can do, for example, \emph{variational}, \emph{integral} or \emph{differential} calculus on a continuous interval of time, a discrete interval of time or a mixing between both of them. The formalism developed in this paper, called \emph{scale dynamics}, is exactly the feature which comes to supplement the time-scale calculus. Indeed, the time-scale calculus only deal with \emph{one scale} whereas, our new objects allows dealing with \emph{multiple scale}. Also, contrary the previous work concerning the change of hypothesis, scale dynamics allows having multiple hypothesis which are encoded in what we called \emph{scale laws}. It shows why there exist different kind of models of the same equation governing the phenomena observed. \\

\textbf{The paper is organized as follows:} \\

In Section 2, we define \emph{scale functions} and \emph{multiscale functions} using the Okamoto's functions as an illustration. We also define a topology which gives a way to analyse the structure of these functions with the help of \emph{symbolic dynamics}. In Section 3, we discuss the modeling problem. Precisely, we come back to the questions of the introduction in order to show the necessity of an analysis of multiscale objects. In Section 4, we introduce the \emph{scale calculus} using the definitions and notations of the \emph{time-scale calculus}. Notably, we define the notion of \emph{scale equation}, \emph{scale laws}, \emph{scale range}, \emph{scale regime} and \emph{scale invariance}. In Section 5, we introduce the \emph{scale dynamics}. Using the scale and multiscale version of Okamoto's functions, we show the first implication of \emph{scale dynamics} on these functions. Then, we provide the general transformation formulas to link the scale structure together. Precisely, we give the formula to quantify the dynamical effects induce by change of scales over the \emph{scale} derivatives. In Section 6 and Section 7, we define \emph{asymptotic differential operators} and \emph{asymptotic scale models} which are illustrated with the \emph{linear} and \emph{fractional} scale regime. This is done using the particular way of scale dynamics to decompose a scale or multiscale function as a ``regular'' and ``irregular'' part. These operators depend on the scale regime and the scale range chosen to construct the asymptotic continuous model. In Section 8, we apply the formalism of scale dynamics to partial differential equations. Using the \emph{scale Newton's equation} we obtain the \emph{diffusion equation} and the \emph{Schr\"odinger equation} under a change of variable. We can see that the \emph{diffusion} is governed by the special scale regime chosen which is the \emph{fractional} scale regime. 

\section{Multiscale functions}

\subsection{Okamoto's functions}

First we remind the classical ``one'' scale Okamoto's function describe in \cite{okamoto}. \\

Let $F_{a}$ be defined inductively over $[0,1]$ by iterations $f_{i}$ for $i\geq 0$ as follows: $f_{0}(x)=x$ for all $x\in[0,1]$, every $f_{i}$ is continuous on $[0,1]$, every $f_{i}$ is affine in each subinterval $[k/3^{i}, (k+1)/3^{i}]$ where $k\in\{0,1,2, \ldots, 3^{i}-1\}$, and
\begin{subequations}
\begin{eqnarray}
f_{i+1}\left(\frac{k}{3^{i}}\right)&=& f_{i}\left(\frac{k}{3^{i}}\right),\label{of1}\\
f_{i+1}\left(\frac{3k+1}{3^{i+1}}\right)&=& f_{i}\left(\frac{k}{3^{i}}\right)+a\left[f_{i}\left(\frac{k+1}{3^{i}}\right)-f_{i}\left(\frac{k}{3^{i}}\right)\right],\label{of2}\\
f_{i+1}\left(\frac{3k+2}{3^{i+1}}\right)&=& f_{i}\left(\frac{k}{3^{i}}\right)+(1-a)\left[f_{i}\left(\frac{k+1}{3^{i}}\right)-f_{i}\left(\frac{k}{3^{i}}\right)\right],\label{of3}\\
f_{i+1}\left(\frac{k+1}{3^{i}}\right)&=& f_{i}\left(\frac{k+1}{3^{i}}\right).\label{of4}
\end{eqnarray}
\end{subequations}

\begin{figure}[h!]
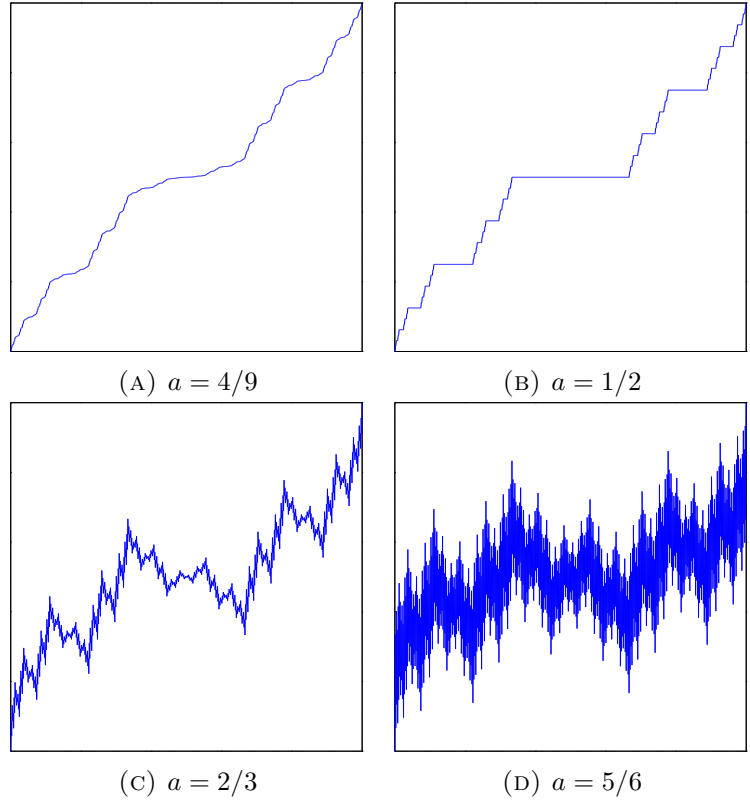

\centering
\begin{subfigure}[b]{0.3\textwidth}
\centering
\resizebox{\textwidth}{!}{\input{lambda49.txt}}
\caption{$a=4/9$}
\end{subfigure}
~
\begin{subfigure}[b]{0.3\textwidth}
\centering
\resizebox{\textwidth}{!}{\input{lambda12.txt}}
\caption{$a=1/2$}
\end{subfigure}
                		
\begin{subfigure}[b]{0.3\textwidth}
\centering
\resizebox{\textwidth}{!}{\input{lambda23.txt}}
\caption{$a=2/3$}
\end{subfigure}
~                		
\begin{subfigure}[b]{0.3\textwidth}
\centering
\resizebox{\textwidth}{!}{\input{lambda56.txt}}
\caption{$a=5/6$}
\end{subfigure}
\caption{Examples of Okamoto's functions}
\label{okamoto_exemples}
\end{figure}

Given this construction, we denote the limit function $F_a$ defined by $\di F_{a}=\lim_{i\rightarrow\infty}f_{i}$. We illustrate the construction with multiple $a$ in Figure \ref{okamoto_exemples}.\\

We have (see \cite[Theorem 1 and Corollary 1, p.2-3]{mccollum}) :

\begin{theorem}
\label{thm_derivability_okamoto}
Let $a\in[0,1]$ and $a_0$ is the unique solution of $54a^3-27a^2=1$. Okamoto's functions have the following properties :

\begin{enumerate}
\item If $a\in[a_{0},1[$, then $F_{a}'(x)$ diverges for almost all $x\in[0,1]$,

\item If $a\in]0,1/3[\cup]1/3, a_{0}[$, then $F_{a}'(x)=0$ for almost all $x\in[0,1]$,

\item If $a=1/3$, then $F_{a}'(x)=1$ for all $x\in[0,1]$.
\end{enumerate}
\end{theorem}
This result is based on the work of \cite{okamoto} and \cite{kobayashi}. \\

The next sections introduce a set of notions including the Okamoto's functions in a wide class of objects called \emph{scale functions} which encode all the classical construction of \emph{fractal functions}. We also define some useful algebraic manipulations on scale functions called \emph{scale composition}.

\subsection{Scale functions}

The previous construction leads to the following objects : 

\begin{itemize}
	\item {\bf Time-scale} : Let $a, b\in \R$, $a<b$. A (discrete and finite) time-scale on $[a,b]$ denoted by $\TT$ is the data of a finite number of points $t_i \in [a,b]$. We denote $\TT =\{ t_i \}$.
	
	\item {\bf Discrete function} : A discrete function is an element of $C(\TT ,\R)$ where $\TT$ is a given time-scale.
	
	\item {\bf PL-Continuous representation of a discrete function} : We denote by $F_{PL}$ the linear interpolation of the discrete function where PL stands for piecewise linear continuous functions.
\end{itemize}
	
In this article, we are concerned with more complex objects depending on scale. In order to make precise this dependence, we introduce the notion of \emph{scale sequences} and \emph{scale functions}.

\begin{definition}[Scale sequence]
Let $a, b\in \R$, $a<b$. A scale sequence denoted by $\PTT$ is the data of a one parameter family of (discrete and finite) time-scale $\TT_i \in [a,b]$ such that $\TT_i \subset \TT_{i+1}$.
\end{definition}

\begin{definition}[Scale function]
Let $\PTT$ be a scale sequence. A scale function is the one parameter family of discrete functions $F_i \in C(\TT_i ,\R)$ denoted by $\mathbf{F}$ such that $F_{i+1} \mid_{\TT_i} =F_i$.
\end{definition}
As an example, the construction of any Okamoto's function induces a scale function. \\

\begin{figure}
	\centering
	\label{figure2_3_0}
	\caption{Scale sequence for Okamoto's scale function}
\end{figure}

\begin{figure}
	\centering
	\resizebox{0.5\textwidth}{!}{\input{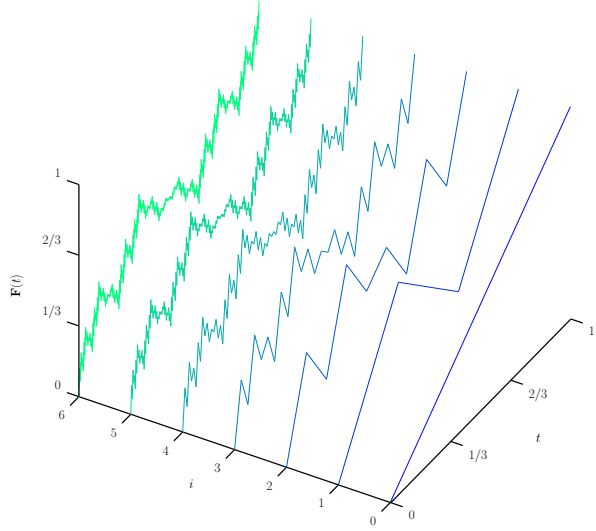}}
	\caption{PL-continuous representation of Okamoto's scale function for $a=\frac{2}{3}$}
	\label{figure2_2}
\end{figure}

A three dimensional representation of scale functions can be obtained as follows : for each $i \in \N^*$, we plot on Figure \ref{figure2_2} the graph of the PL-continuous representation of $F_i$ denotes by $\Gamma_i$.

\subsection{Scale elementary action}

In order to define a scale composition, we introduce the notion of \emph{elementary time-scale} which are time-scales reduced to only two points, i.e. $\TT_{elem }$ is always of the form $\TT_{elem} =\{ t_0 ,t_1\}$. Any (discrete and finite) time-scale can be decomposed in a union of elementary time-scales as follows :\\

Let $\TT =\{ t_ 0 ,\dots ,t_n \}$. We denote $\TT_{elem,i} =\{ t_i ,t_{i+1} \}$ for $i= 0,\dots ,n-1$. We have 
\begin{equation}
\TT =\bigcup_{i=1}^{n-1} \TT_{elem,i} .
\end{equation}
We denote by $\mbox{Elem}$ the operator acting on (discrete and finite) time-scale producing the decomposition in elementary time-scales, i.e. 
\begin{equation}
\mbox{Elem} (\TT ) =\left \{ \TT_{elem,i} \right \}_{i=0 ,\dots ,n-1} .
\end{equation}

An \emph{elementary discrete function} is a discrete function defined on an elementary time-scale. These discrete functions are the basic piece to describe a given elementary pattern. Precisely, we have :

\begin{definition}[Scale elementary action]
A \emph{scale elementary action} is the data of an operator $A : C(\TT_{elem} ,\R ) \rightarrow C(\TT_{elem,A} ,\R )$ satisfying the following properties : 
\begin{itemize}
	\item $\TT_{elem,A}$ is such that $\TT_{elem} \subset \TT_{elem,A}$ and where the intervals associated with the time-scale are such that $I_{\TT_{elem},A} \subset I_{\TT_{elem}}$.
	\item For all $F \in C(\TT_{elem},\R)$, the action of $A$ on $F$ denoted by $A\circledcirc F$ with $A\circledcirc F \in C(\TT_{elem,A} ,\R)$ satisfies 
	\begin{equation}
	(A \circledcirc F )\mid_{\TT_{elem}} =F .
	\end{equation}	
	\end{itemize}
\end{definition}

An example of scale elementary action is given by the Okamoto's construction. 

\begin{definition}[Scale elementary Okamoto action]
\label{def_scale_elem_okamoto}
Let $\TT_{elem} =\{ t_0 ,t_1 \}$ and $a\in ]0,1[$. We denote by $O_a$ and we call the \emph{scale elementary Okamoto action} the scale elementary action defined by :
\begin{itemize}
\item The time-scale $\TT_{elem,O_a}$ : 
\begin{equation}
\TT_{elem,O_a} =\{ t_0 , t_{0,1} , t_{0,2} ,t_1 \} ,
\end{equation}
where $t_{0,i} =t_0 +i \di \frac{\mu}{3} ,\ i=1,2,\ \mu =t_1 -t_0$.

\item The action $O_a$ : for $F\in C(\TT_{elem},\R)$ we have $O_a (t_0) =F (t_0 )$, $O_a (t_1 )=F (t_1)$ and
\begin{equation}
O_a (t_{0,1}) = F(t_0) + a (F (t_1) -F(t_0)) ,\ 
O_a (t_{0,2}) = F(t_0) + (1-a) (F (t_1) -F(t_0)) .
\end{equation}
\end{itemize}
\end{definition}

In order to see the \emph{concrete} action of $O_a$ to produce an elementary pattern, we apply $O_a$ on the elementary discrete function $E_0 : \TT_0 =\{ 0,1 \} \rightarrow \R$ defined by $E_0 (0)=0$ and $E(1)=1$. We obtain the following picture for $O_a \circledcirc E_0 \in C(\TT_1 ,\R )$ :

\begin{figure}[h!]
\centering
	\begin{subfigure}[b]{0.45\textwidth}
	\centering
	\resizebox{\textwidth}{!}{
\begin{pgfpicture}
\pgfsetlinewidth{0.01pt}
\color[rgb]{1.000000,1.000000,1.000000}
\pgfpathmoveto{\pgfpoint{130.000031pt}{893.614990pt}}
\pgflineto{\pgfpoint{905.000061pt}{118.614990pt}}
\pgflineto{\pgfpoint{130.000031pt}{118.614990pt}}
\pgfpathclose
\pgfusepath{fill,stroke}
\pgfpathmoveto{\pgfpoint{130.000031pt}{893.614990pt}}
\pgflineto{\pgfpoint{905.000061pt}{893.614990pt}}
\pgflineto{\pgfpoint{905.000061pt}{118.614990pt}}
\pgfpathclose
\pgfusepath{fill,stroke}
\color[rgb]{0.000000,0.000000,0.000000}
\pgfsetlinewidth{2.000000pt}
\pgfsetdash{{16pt}{0pt}}{0pt}
\pgfpathmoveto{\pgfpoint{905.000061pt}{118.614990pt}}
\pgflineto{\pgfpoint{130.000031pt}{118.614990pt}}
\pgfusepath{stroke}
\pgfpathmoveto{\pgfpoint{905.000061pt}{893.614990pt}}
\pgflineto{\pgfpoint{130.000031pt}{893.614990pt}}
\pgfusepath{stroke}
\pgfpathmoveto{\pgfpoint{905.000061pt}{893.614990pt}}
\pgflineto{\pgfpoint{130.000031pt}{893.614990pt}}
\pgfusepath{stroke}
\pgfpathmoveto{\pgfpoint{905.000061pt}{118.614990pt}}
\pgflineto{\pgfpoint{130.000031pt}{118.614990pt}}
\pgfusepath{stroke}
\pgfpathmoveto{\pgfpoint{130.000031pt}{893.614990pt}}
\pgflineto{\pgfpoint{130.000031pt}{118.614990pt}}
\pgfusepath{stroke}
\pgfpathmoveto{\pgfpoint{905.000061pt}{893.614990pt}}
\pgflineto{\pgfpoint{905.000061pt}{118.614990pt}}
\pgfusepath{stroke}
\pgfpathmoveto{\pgfpoint{905.000061pt}{893.614990pt}}
\pgflineto{\pgfpoint{905.000061pt}{118.614990pt}}
\pgfusepath{stroke}
\pgfpathmoveto{\pgfpoint{130.000031pt}{893.614990pt}}
\pgflineto{\pgfpoint{130.000031pt}{118.614990pt}}
\pgfusepath{stroke}
\pgfpathmoveto{\pgfpoint{130.000031pt}{893.614990pt}}
\pgflineto{\pgfpoint{130.000031pt}{893.614990pt}}
\pgfusepath{stroke}
\pgfpathmoveto{\pgfpoint{905.000061pt}{893.614990pt}}
\pgflineto{\pgfpoint{905.000061pt}{893.614990pt}}
\pgfusepath{stroke}
\pgfpathmoveto{\pgfpoint{905.000061pt}{118.614990pt}}
\pgflineto{\pgfpoint{905.000061pt}{118.614990pt}}
\pgfusepath{stroke}
\pgfpathmoveto{\pgfpoint{130.000031pt}{118.614990pt}}
\pgflineto{\pgfpoint{130.000031pt}{118.614990pt}}
\pgfusepath{stroke}
\pgfpathmoveto{\pgfpoint{130.000031pt}{128.578369pt}}
\pgflineto{\pgfpoint{130.000031pt}{118.614990pt}}
\pgfusepath{stroke}
\pgfpathmoveto{\pgfpoint{130.000031pt}{883.651611pt}}
\pgflineto{\pgfpoint{130.000031pt}{893.614990pt}}
\pgfusepath{stroke}
\pgfpathmoveto{\pgfpoint{905.000061pt}{128.578369pt}}
\pgflineto{\pgfpoint{905.000061pt}{118.614990pt}}
\pgfusepath{stroke}
\pgfpathmoveto{\pgfpoint{905.000061pt}{883.651611pt}}
\pgflineto{\pgfpoint{905.000061pt}{893.614990pt}}
\pgfusepath{stroke}
{
\pgftransformshift{\pgfpoint{130.000031pt}{113.615051pt}}
\pgfnode{rectangle}{north}{\fontsize{20}{0}\selectfont\textcolor[rgb]{0,0,0}{{$0$}}}{}{\pgfusepath{discard}}}
{
\pgftransformshift{\pgfpoint{905.000061pt}{113.615051pt}}
\pgfnode{rectangle}{north}{\fontsize{20}{0}\selectfont\textcolor[rgb]{0,0,0}{{$1$}}}{}{\pgfusepath{discard}}}
\pgfpathmoveto{\pgfpoint{139.963379pt}{118.614990pt}}
\pgflineto{\pgfpoint{130.000031pt}{118.614990pt}}
\pgfusepath{stroke}
\pgfpathmoveto{\pgfpoint{895.036743pt}{118.614990pt}}
\pgflineto{\pgfpoint{905.000061pt}{118.614990pt}}
\pgfusepath{stroke}
\pgfpathmoveto{\pgfpoint{139.963379pt}{893.614990pt}}
\pgflineto{\pgfpoint{130.000031pt}{893.614990pt}}
\pgfusepath{stroke}
\pgfpathmoveto{\pgfpoint{895.036743pt}{893.614990pt}}
\pgflineto{\pgfpoint{905.000061pt}{893.614990pt}}
\pgfusepath{stroke}
{
\pgftransformshift{\pgfpoint{125.000031pt}{118.614990pt}}
\pgfnode{rectangle}{east}{\fontsize{20}{0}\selectfont\textcolor[rgb]{0,0,0}{{$0$}}}{}{\pgfusepath{discard}}}
{
\pgftransformshift{\pgfpoint{125.000031pt}{893.614990pt}}
\pgfnode{rectangle}{east}{\fontsize{20}{0}\selectfont\textcolor[rgb]{0,0,0}{{$1$}}}{}{\pgfusepath{discard}}}
{
\pgftransformshift{\pgfpoint{517.500061pt}{89.614990pt}}
\pgfnode{rectangle}{north}{\fontsize{20}{0}\selectfont\textcolor[rgb]{0,0,0}{{$t$}}}{}{\pgfusepath{discard}}}
\color[rgb]{0.000000,0.000000,1.000000}
\pgfsetdash{}{0pt}
\pgfpathmoveto{\pgfpoint{905.000061pt}{893.614990pt}}
\pgflineto{\pgfpoint{130.000031pt}{118.614990pt}}
\pgfusepath{stroke}
\pgfsetlinewidth{0.01pt}
\pgfpathmoveto{\pgfpoint{135.000000pt}{118.614990pt}}
\pgflineto{\pgfpoint{131.545074pt}{113.859741pt}}
\pgflineto{\pgfpoint{134.045105pt}{115.676086pt}}
\pgfpathclose
\pgfusepath{fill,stroke}
\pgfpathmoveto{\pgfpoint{135.000000pt}{118.614990pt}}
\pgflineto{\pgfpoint{128.454926pt}{113.859741pt}}
\pgflineto{\pgfpoint{131.545074pt}{113.859741pt}}
\pgfpathclose
\pgfusepath{fill,stroke}
\pgfpathmoveto{\pgfpoint{135.000000pt}{118.614990pt}}
\pgflineto{\pgfpoint{125.954895pt}{115.676086pt}}
\pgflineto{\pgfpoint{128.454926pt}{113.859741pt}}
\pgfpathclose
\pgfusepath{fill,stroke}
\pgfpathmoveto{\pgfpoint{135.000000pt}{118.614990pt}}
\pgflineto{\pgfpoint{125.000000pt}{118.614990pt}}
\pgflineto{\pgfpoint{125.954895pt}{115.676086pt}}
\pgfpathclose
\pgfusepath{fill,stroke}
\pgfpathmoveto{\pgfpoint{135.000000pt}{118.614990pt}}
\pgflineto{\pgfpoint{125.954895pt}{121.553955pt}}
\pgflineto{\pgfpoint{125.000000pt}{118.614990pt}}
\pgfpathclose
\pgfusepath{fill,stroke}
\pgfpathmoveto{\pgfpoint{135.000000pt}{118.614990pt}}
\pgflineto{\pgfpoint{128.454926pt}{123.370239pt}}
\pgflineto{\pgfpoint{125.954895pt}{121.553955pt}}
\pgfpathclose
\pgfusepath{fill,stroke}
\pgfpathmoveto{\pgfpoint{135.000000pt}{118.614990pt}}
\pgflineto{\pgfpoint{131.545074pt}{123.370239pt}}
\pgflineto{\pgfpoint{128.454926pt}{123.370239pt}}
\pgfpathclose
\pgfusepath{fill,stroke}
\pgfpathmoveto{\pgfpoint{135.000000pt}{118.614990pt}}
\pgflineto{\pgfpoint{134.045105pt}{121.553955pt}}
\pgflineto{\pgfpoint{131.545074pt}{123.370239pt}}
\pgfpathclose
\pgfusepath{fill,stroke}
\pgfpathmoveto{\pgfpoint{910.000000pt}{893.614990pt}}
\pgflineto{\pgfpoint{906.545105pt}{888.859741pt}}
\pgflineto{\pgfpoint{909.045166pt}{890.676086pt}}
\pgfpathclose
\pgfusepath{fill,stroke}
\pgfpathmoveto{\pgfpoint{910.000000pt}{893.614990pt}}
\pgflineto{\pgfpoint{903.454956pt}{888.859741pt}}
\pgflineto{\pgfpoint{906.545105pt}{888.859741pt}}
\pgfpathclose
\pgfusepath{fill,stroke}
\pgfpathmoveto{\pgfpoint{910.000000pt}{893.614990pt}}
\pgflineto{\pgfpoint{900.954956pt}{890.676086pt}}
\pgflineto{\pgfpoint{903.454956pt}{888.859741pt}}
\pgfpathclose
\pgfusepath{fill,stroke}
\pgfpathmoveto{\pgfpoint{910.000000pt}{893.614990pt}}
\pgflineto{\pgfpoint{900.000000pt}{893.614990pt}}
\pgflineto{\pgfpoint{900.954956pt}{890.676086pt}}
\pgfpathclose
\pgfusepath{fill,stroke}
\pgfpathmoveto{\pgfpoint{910.000000pt}{893.614990pt}}
\pgflineto{\pgfpoint{900.954956pt}{896.553955pt}}
\pgflineto{\pgfpoint{900.000000pt}{893.614990pt}}
\pgfpathclose
\pgfusepath{fill,stroke}
\pgfpathmoveto{\pgfpoint{910.000000pt}{893.614990pt}}
\pgflineto{\pgfpoint{903.454956pt}{898.370239pt}}
\pgflineto{\pgfpoint{900.954956pt}{896.553955pt}}
\pgfpathclose
\pgfusepath{fill,stroke}
\pgfpathmoveto{\pgfpoint{910.000000pt}{893.614990pt}}
\pgflineto{\pgfpoint{906.545105pt}{898.370239pt}}
\pgflineto{\pgfpoint{903.454956pt}{898.370239pt}}
\pgfpathclose
\pgfusepath{fill,stroke}
\pgfpathmoveto{\pgfpoint{910.000000pt}{893.614990pt}}
\pgflineto{\pgfpoint{909.045166pt}{896.553955pt}}
\pgflineto{\pgfpoint{906.545105pt}{898.370239pt}}
\pgfpathclose
\pgfusepath{fill,stroke}
\end{pgfpicture}}
	\caption{$E_0$}
	\end{subfigure}
	~                		
	\begin{subfigure}[b]{0.465\textwidth}
	\centering
	\resizebox{\textwidth}{!}{\input{figure2_3_2.txt}}
	\caption{$O_a \circledcirc E_0$}
	\end{subfigure}
\end{figure}

\subsection{Scale action on discrete functions}

This Section is devoted to the definition of a scale action which is the formal formulation of the classical idea of \emph{iterative construction} of fractal function using a given \emph{elementary pattern}. The elementary pattern is encoded by the scale elementary action. The new figure obtained by an iteration of this pattern on a given discrete function will be encoded by the \emph{scale action}. Precisely, we have :

\begin{definition}[Scale action]
Let $A$ be an elementary scale action. Let  $F\in C(\TT ,\R)$ where $\TT$ is an arbitrary (discrete and finite) time-scale.  

\begin{itemize}
\item Let ${\rm Elem}(\TT) =\{ \TT_i\}_{i=0,\dots ,n-1}$. For each $\TT_i \in {\rm Elem}(\TT)$, as $F\mid_{\TT_i} \in C(\TT_i ,\R)$ is an elementary function, the discrete function $A(F\mid_{\TT_i}) \in C(\TT_{i,A} ,\R)$ is well defined. We denote by $\TT_A$ the time-scale 
\begin{equation}
\TT_A=\bigcup_{i=0}^{n-1}\TT_{i,A} .
\end{equation}
\end{itemize}

The scale action induced by $A$ on $F$ denoted by $A\circledcirc F \in C(\TT_A ,\R)$ is defined by 
\begin{equation}
\left [ A\circledcirc F \right ]\mid_{\TT_{i,A}}=A \circledcirc \left [ F\mid_{\TT_i} \right ]
\end{equation}
for $i=0,\dots ,n-1$.
\end{definition}

The scale action associated to an elementary scale action $A$ allows us to make composition of a given operator on a discrete function. Let $A$ be an elementary action. We denote by $A^{\circledcirc n}$ the operator defined over discrete function given by the composition of the action of $A$ $n$ times.\\

For example, let us consider the iterative action of the scale Okamoto elementary action on the discrete function $E_0$. We denote by $E_i =O_a^{\circledcirc i} (E_0 ) \in C(\TT_i ,\R)$ where $\TT_i =\TT_{i-1 ,O}$ with $T_0 =\{ 0,1\}$. The discrete function $E_2$ looks like
\begin{figure}[h!]
	\centering
	\resizebox{0.45\textwidth}{!}{\input{figure2_3_3.txt}}
	\caption{$E_2=O_a \circledcirc E_1 = O_a^{\circledcirc 2}(E_0)$}
\end{figure}

\subsection{Multiscale functions}

In this Section we introduce a toy model to explain the difficulties related to the modeling of complex phenomenon from a finite set of data. More complex constructions can be made and we discuss some of them in the next Section.

\begin{definition}[Multiscale functions]
	Let $\mathbf{A} =\{ A_i \}$ be a sequence (potentially infinite) of scale elementary actions and $\mathbf{N}=(N_1 ,\dots ,N_n)$ such that $N_i \in \mathbb{N}^*$, $i=1,\dots ,n$ be the complexity pattern. If $n$ is finite then $N_n =\infty$ otherwise $N_i$ are all finite. The multiscale function of order $m$ associated to $\mathbf{A}$ and $\mathbf{N}$ is the function defined by
	\begin{equation}
	F_{\mathbf{A} ,\mathbf{N},m} =A_{k}^{\circledcirc N_k^m} \circledcirc A_{k-1}^{\circledcirc N_{k-1}} \circledcirc \dots \circledcirc A_{1}^{\circledcirc N_1} \circledcirc E_0,
	\end{equation}
	where $k$ satisfies $N_1 + \dots + N_{k-1} \leq m$ and $N_1 + \dots + N_k \geq m$ and $N_k^m =m-(N_1 +\dots +N_{k-1})$. 
	
	We denote by $F_{\mathbf{A} ,\mathbf{N}}$ the limit of this function when $m$ goes to infinity. Such a function is called a multiscale function.
\end{definition}

The three sequences $\mathbf{A}$, $\mathbf{N}$ and $\PTT$ encodes the \emph{scale structure} of a given multiscale function.\\

As an example, we can define the Multiscale Okamoto's functions as follows : 

\begin{definition}[Multiscale Okamoto's functions]
\label{def_multiscale_okamoto}
Let $\mathbf{a}=(a_1 ,\dots ,a_n )$ be a sequence (potentially infinite) $a_i\in ]0,1[$ and $\mathbf{N}=(N_1 ,\dots ,N_n)$ such that $N_i \in \mathbb{N}^*$, $i=1,\dots ,n$. If $n$ is finite then $N_n =\infty$ otherwise $N_i$ are all finite. The multiscale Okamoto's function of order $m$ is the function defined by
\begin{equation}
O_{\mathbf{a} ,\mathbf{N},m} =O_{a_k}^{\circledcirc N_k^m} \circledcirc O_{a_{k-1}}^{\circledcirc N_{k-1}} \circledcirc \dots \circledcirc O_{a_1}^{\circledcirc N_1} \circledcirc E_0,
\end{equation}
where $k$ satisfies $N_1 + \dots + N_{k-1} \leq m$ and $N_1 + \dots + N_k \geq m$ and $N_k^m =m-(N_1 +\dots +N_{k-1})$. 

We denote by $O_{\mathbf{a} ,\mathbf{N}}$ the limit of this function when $m$ goes to infinity. Such a function is called a multiscale Okamoto's functions.
\end{definition}

The three sequences $\mathbf{a}$, $\mathbf{N}$ and $\PTT$ encodes the \emph{scale structure} of a given multiscale Okamoto function.\\

We denote by $\mathbf{Okamoto}$ the set of all multiscale Okamoto's functions. In what follows, we abbreviate a multiscale Okamoto's function by \emph{MSO function}.

\subsection{Structure of MSO functions : A characterization using symbol sequences}
In order to characterize multiscale Okamoto's function with symbol sequences, we use some well known construction on space of symbol sequences as exposed in \cite[$\S$.2.2.a p. 96]{wig}.

\begin{definition}[Space of symbol sequences]
	Let $A$ be a subset of a given metric space. We denote by $\Sigma_A$ the space of symbol sequences on $A$ defined by an infinite Cartesian product of copies of $A$, i.e.
	\begin{equation}
	\di\Sigma_A =\prod_{i=1}^{\infty} A .
	\end{equation}
	An element of $\Sigma_A$ is called an infinite sequence or word defined on $A$ and is denotes by
	\begin{equation}
	\mathbf{s}\in \Sigma_A,\ \ \mathbf{s}=(s_1 ,\dots ,s_n ,\dots )\ \ \mbox{where}\ s_i \in A, \ \forall i \in \N .
	\end{equation}
\end{definition}

Let $\mathbf{a},\mathbf{N}$ be two admissible sequences. We denote by $s_{\mathbf{a},\mathbf{N}}$ the infinite sequence in $\Sigma_{[0,1]}$ defined by 
\begin{equation}
\label{sequence_multiscale}
s_{\mathbf{a},\mathbf{N}} =(a_1 \ldots a_1 a_2 \ldots a_2 \ldots )
\end{equation}
where $a_1$ is repeated $N_1$ times, $s_2$ is repeated $N_2$ times and so on. Using this construction, we have the following Lemma :
\begin{lemma}
	The set $\mathbf{Okamoto}$ is in one to one correspondence with $\Sigma_{[0,1]}$.
\end{lemma}

\subsubsection{Topology on $\mathbf{Okamoto}$}

We introduce a topology on the set of multiscale Okamoto's functions using the previous characterization.\\ 

A topology on $\Sigma_A$ can be defined as follows.  

\begin{definition}[Metric]
	Let $A$ be a metric space and $d$ its metric. A metric on $\Sigma_A$ is given for all $\mathbf{s},\mathbf{s}' \in \Sigma_A$ by 
	\begin{equation}
	d_{\Sigma} (\mathbf{s},\mathbf{s}') =\di\sum_{i=1}^{\infty} \di\frac{1}{2^i} \di\frac{d(s_i ,s_i')}{1+d(s_i ,s_i')} .
	\end{equation}	
\end{definition}

We refer to \cite[p.98]{wig}.\\

The previous metric induces a topology on the set of multiscale Okamoto's functions via the one-to-one correspondence with $\Sigma_{[0,1]} \times \Sigma_{\N^*}$ as follows :

\begin{definition}
	A metric on $\mathbf{Okamoto}$ is given for all $f_{\mathbf{a},\mathbf{N}}$ and $f_{\mathbf{a}',\mathbf{N}'} \in \mathbf{Okamoto}$ by 
	\begin{equation}
	d_{\mathbf{Okamoto}} (f_{\mathbf{a},\mathbf{N}} ,f_{\mathbf{a}',\mathbf{N}'}) = 
	\di d_{\Sigma_{[0,1]}} (s_{\mathbf{a} ,\mathbf{N}},s_{\mathbf{a}' ,\mathbf{N}'}) .
	\end{equation}	
\end{definition}

\subsubsection{Symbolic dynamics and the three basic classes}

The study of multiscale Okamoto's functions is related to the behavior of the curve under change of scales. Precisely, the data of the two sequences $\mathbf{a}$ and $\mathbf{N}$ encodes the way the curve behaves over the set of scales $\PTT=\{ \TT_i \}_{i\in \N*}$. The change from scale $\TT_i$ to $\TT_{i+1}$ is then associated on the sequences side to the classical \emph{shift map} $\sigma : \Sigma_A \rightarrow \Sigma_A$ defined for all $\mathbf{s} \in \Sigma_A$ by \cite[$\S$.2.2.b p.100]{wig} :
\begin{equation}
\sigma (\mathbf{s})_i =\sigma_{i+1} , \ i\in \N^* .
\end{equation}
The shift map is the simplest example of a \emph{chaotic map} (see \cite[$\S$.2.1.e p.93]{wig}). The main property of these maps is summarized in the following Theorem (see \cite[Prop.2.2.11 p.105]{wig}):

\begin{theorem}
	The shift map possesses the following properties :
	\begin{enumerate}
		\item a countable infinity of periodic orbits,
		\item an uncountable infinity of non periodic orbits, 
		\item a dense orbit.
	\end{enumerate}
\end{theorem}
This Theorem can be used to classify Multiscale Okamoto functions depending on their Scale structure. Precisely, we introduce the following classes of functions :
\begin{itemize}
	\item Self similar MSO functions denoted $\mathbf{Okamoto}_{self}$ corresponding to periodic sequences.
	\item Random MSO functions denoted by $\mathbf{Okamoto}_{rand}$ corresponding to non periodic sequences.
	\item The Chaotic MSO function corresponding to the dense orbit.
\end{itemize}
It must be noted that the scale structure of a given MSO function can be very complicated independently of its shape structure which is more related to its regularity. It means that the previous encoding does not capture the geometrical complexity of a MSO function. This is precisely the starting point of the \emph{scale dynamics} studied in the following Sections.

%

\subsubsection{Complexity of Multiscale Okamoto's functions}

There exists several notions of complexity related to symbol sequences in \emph{combinatoric}. Using these notions, it is possible to select from a given symbol sequence a natural candidate with low complexity and representing the associated multiscale Okamoto's function in one of the previous family. We refer to the report of J-P. Allouche in \cite{al} for more details and precise definition of these notions.

\subsection{Possible generalizations ?}

The previous definitions and constructions can of course be generalized in many directions. For example, the actual definition of multiscale functions use the same elementary action in order to produce the function between two successive scales. In order to define more complicated objects, one can allow a given set of elementary actions to go from one scale to the next one, producing mixed multiscale structures on a given function. The formalization of this kind of objects is of course a little complicated but does not bring new fundamental ideas despite its interest. As a consequence, we prefer to discuss the construction of our formalism in the limited (but already substantial) setting of multiscale functions.

\section{The modeling problem}

In the following, a multiscale Okamoto's function is the mathematical analogue of a real physical process. It means that we assume that the exact behavior of the system is given by such a function. The modeling problem is then formulated as follows : \\

{\bf Modeling problem}: {\it Having a finite set of data concerning a given multiscale Okamoto's function, can we recover the nature of the underlying function ?}\\

As we have already discussed in the first Section, the nature of an object is for example : random or not ? or depending on scales ? etc.\\

As we will see in the following, the answer is \emph{no}. Only an infinite amount of data can give a complete characterization of the nature of an object. This result has strong 
consequences, as it means that no models can decide the exact nature of a physical phenomenon. In order to do so, we need some extra conditions which are coming from physics and more precisely from the underlying framework of a given theory. As an example in general relativity, the models are constructed assuming that no randomness occurs and that all the objects can be described in the differential framework. In other words, an identification of a given ``exact'' model is related to a given ``philosophy of nature''. \\

The previous modeling problem is not sufficiently precise in order to formulate a result. 
Using these results, we can precise the modeling problem for multiscale Okamoto's functions.\\

{\bf Modeling problem 2}. {\it Let $m\in \N^*$ be given and $f_{\mathbf{a},\mathbf{N}} \in \mathbf{Okamoto}$. Can we decide the ``nature'' of $f_{\mathbf{a},\mathbf{N}}$ knowing the associated multiscale Okamoto's function of order $m$ for arbitrary $m$ ?}

\subsection{Toward scale dynamics}

The previous discussion proves that the complexity of a given multiscale function is not related to the nature of the limit function. The basic information which is missing is the behavior of the \emph{scale derivatives} (left and right) during the scale process, i.e. one needs to introduce a \emph{scale dynamics}. This is precisely what is developed in the next Section.

\section{Scale calculus}

In this Section, we define the notion of \emph{scale derivative} which will be used in the next Section in order to define what we call a \emph{scale dynamics}. Our definition is based on some classical tools of \emph{discrete calculus} (finite differences). We use the notations which are usual in the theory of the \emph{time-scale calculus} indicating by this way that most of the notion defined in this section can be generalized over multiscale functions with a general time-scale sequence.  

\subsection{Reminder about time-scale calculus}

We consider $\TT$ a discrete and finite time-scale with $a = \min (\TT)$, $b = \max (\TT)$ and $\mathrm{card} (\T) \geq 3$.

\begin{definition}
	The backward and forward jump operators $\rho, \sigma : \TT \longrightarrow \TT$ are respectively defined for all $\forall t \in \TT$ by :
	\begin{equation*}
	\rho (t) = \sup \{ s \in \TT, \; s < t \} \; \text{and} \; \sigma (t) = \inf \{ s \in \TT, \; s > t \},
	\end{equation*}
	where we put $\sup \emptyset = a$ and $\inf \emptyset = b$.
\end{definition}

\begin{definition}
	The forward graininess (resp. backward graininess) function $\fonctionsansdef{\mu}{\T}{\R^+}$ (resp. $\fonctionsansdef{\nu}{\T}{\R^+}$) is defined by $\mu(t) = \sigma (t) -t$ (resp. $\nu(t) = t- \rho (t)$) for any $t \in \T$.
\end{definition}

We set $\TK = \T \backslash ]\rho(b),b]$, $\Tk = \T \backslash [a,\sigma(a)[$ and $\TKk = \TK \cap \Tk$. Let us recall the usual definitions of $\Delta$- and $\nabla$-differentiability.

\begin{definition}
	A function $\fonctionsansdef{u}{\T}{\R^n}$, where $n \in \N^*$, is said to be $\Delta$-differentiable at $t \in \TK$ (resp. $\nabla$-differentiable at $t \in \Tk$) if the following limit exists in $\R^n$:
	\begin{equation}
	\lim\limits_{\substack{s \to t \\ s \neq \sigma (t) }} \dfrac{u(\sigma(t))-u(s)}{\sigma(t) -s} \; \left( \text{resp.} \; \lim\limits_{\substack{s \to t \\ s \neq \rho (t) }} \dfrac{u(s)-u(\rho (t))}{s-\rho(t)} \right).
	\end{equation}
	In such a case, this limit is denoted by $\Delta u(t)$ (resp. $\nabla u(t)$).
\end{definition}

Let us denote by $\int \Delta \tau$ the Cauchy $\Delta$-integral defined in \cite[p.26]{bohn} with the following result.

\begin{theorem}[{\cite[Theorem 1.74 p.27]{bohn}}]
	For every $u \in \Crd(\TK)$, there exist a unique $\Delta$-antiderivative of $u$ in sense of $\Delta U= u$ on $\TK$ vanishing at $t=a$. In this case the $\Delta$-integral is defined by
	\begin{equation*}
	U(t) = \int_a^t u(\tau) \Delta \tau
	\end{equation*}
	for every $t \in \T$.
\end{theorem}

\subsection{The scale derivative}

The aim of this Section is to define the natural object encoding the behavior of the $\Delta$ or $\nabla$ derivatives over the sequence of time-scales associated to a given scale function.

\begin{definition}[$\Delta$ Scale derivative]
Let $\mathbf{F}$ be a given scale function over the scale sequence $\PTT$. The $\Delta$ scale derivative of $\mathbf{F}$ is the scale function denoted by $\Delta (\mathbf{F})$ and is defined by :
	
\begin{itemize}
	\item ${\TT_{\Delta (\mathbf{F}),i}} =\TT_i^{\kappa}$.
	\item $\left [ \Delta (\mathbf{F}) \right ]_i =\Delta (F_i )\in C(\TT_i^{\kappa} ,\R)$ .
\end{itemize}
\end{definition}

As an example, the scale derivative of the scale Okamoto function of order $a=\frac{2}{3}$ is given by
\begin{figure}[ht!]
\centering
\resizebox{0.5\textwidth}{!}{\input{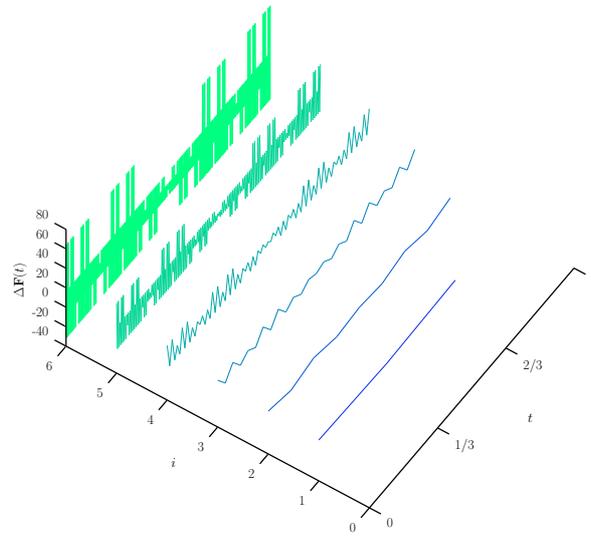}}
\caption{$\Delta$ Scale derivative}
\end{figure}
For multiscale functions, the scale derivative is a priori \emph{not} a multiscale function but \emph{only} a scale function. This remarks put in evidence the disconnection between the complexity in scale of a given function and its complexity from the scale dynamical view point. \\

A notion of $\nabla$ scale derivative is defined in the same way. We denote by $C_\Delta(\PTT)$ (resp. $C_\nabla(\PTT)$) the set of functions which are \emph{scale $\Delta$ differentiable} (resp. \emph{scale $\nabla$ differentiable}) over $\PTT$.

\subsection{The scale antiderivative}
Although we will not use scale antiderivative in a first approach to scale dynamics, we will need for some applications to integrate scale functions. As a consequence, we provide the corresponding notion.

\begin{definition}[$\Delta$ Scale antiderivative]
	Let $\mathbf{F}$ be a given scale function over the scale sequence $\PTT$. The $\Delta$ scale antiderivative of $\mathbf{F}$ is the scale function denoted by $\di \int \mathbf{F}\, \Delta $ and is defined over the scale sequence $\PTT$ by
	\begin{equation}
		\left  [\di \int_{t_0} \mathbf{F} \, \Delta  \right ]_i =\int_{t_0} F_i \, \Delta \in C(\TT_i ,\R) .
	\end{equation}
\end{definition}
As an example, the scale antiderivative of the scale Okamoto function of order $a=\frac{2}{3}$ is given by 
\begin{figure}[ht!]
\centering
\resizebox{0.5\textwidth}{!}{\input{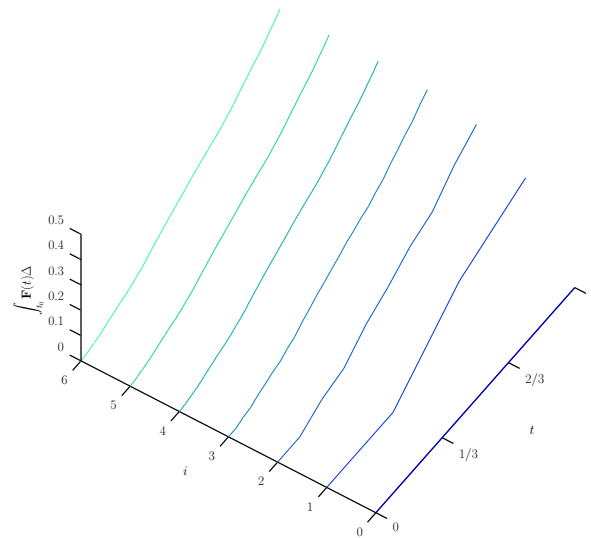}}
\caption{$\Delta$ Scale antiderivative}
\end{figure}

\subsection{Scale equations}

The scale derivative and antiderivative can be used to define what can be called \emph{scale equations} which are more or less time-scale equations on each scales. These equations will have an important role in the last Section concerning applications of scale calculus and scale dynamics to classical mechanics.\\

We define \emph{scale equations} as follows :

\begin{definition}
	[Scale equations] A scale equation is a formal relation written on scale functions $\mathbf{F}$ on a given scale sequence $\PTT$ of the form
	\begin{equation}
	R_{\TT_i} (F_i , \Delta , \di\int \cdot \Delta )=0 ,
	\end{equation}
where $\mathbf{F}=\{ F_i \}$ and $\PTT =\{ \TT_i \}$.
\end{definition}

As an example, we can consider the linear scale equation given by 
\begin{equation}
\label{linearscale}
\Delta (F_i ) - \mu (\TT_i) F_i =0.
\end{equation}

Some applications will lead to special scale equations which we call \emph{scale invariant} and are defined as follows :

\begin{definition}
	[Scale invariance] A scale equation is scale invariant if the scale equation
	\begin{equation}
	R_{\TT_i} \left(F_i , \Delta , \di\int \cdot \Delta \right)=0 ,
	\end{equation}
	where $\mathbf{F}=\{ F_i \}$ and $\PTT =\{ \TT_i \}$ satisfies
	\begin{equation}
	R_{\TT_i} =R ,
	\end{equation}
where $R$ is a fixed relation.
\end{definition}

The previous linear scale equation (\ref{linearscale}) is not scale invariant. Indeed, the scale operator is given by
\begin{equation}
R_{\TT_i} \left(F_i , \Delta , \di\int \cdot \Delta \right)= \Delta (F_i ) - \mu (\TT_i) F_i ,
\end{equation}
which is explicitly scale dependent through the graininess constant $\mu (\TT_i)$ which depends on each scale $\TT_i$. \\
 
As an example of a scale invariant equation, we introduce the scale Euler-Lagrange equation which will be studied in the last Section.\\

A \emph{Scale Euler-Lagrange equation} is defined for all $\mathbf{F}$ by 
\begin{equation}
\nabla \left ( \di \frac{\partial L}{\partial v} (F_i ,\Delta F_i )\right ) = \di\frac{\partial L}{\partial x} (F_i ,\Delta F_i ).
\end{equation}
In this case the relation is given by the \emph{Euler operator} using scale calculus and given by 
\begin{equation}
R_{\mbox{\rm Euler}} = \nabla \circ \frac{\partial L}{\partial v} - \frac{\partial L}{\partial x} ,
\end{equation}
over scale functions whose form is independent of scales.\\

Of course, more complex notion of invariance can be defined. For example, in many physical problems we can expect only a \emph{partial} scale invariance, meaning that the equation keep the same form only on a given range of scales. This problem will be discussed in details in Section \ref{continuous}.

\subsection{Scale equations and scale regimes}

The study of scale equations depends mainly as in the classical theory on the underlying functional space on which such an equation is studied. In the following, in particular in Section \ref{continuous}, we will deal with the dynamical behavior of scale equations over multiscale functions. In that case, the notion of \emph{scale regime} will have a very important role.\\

In order to identify scale regime for a given scale function, one need to select a given class of \emph{scale comparison functions}. Examples are given by the classical \emph{Hardy scale}, the \emph{logarithmic scale} and the \emph{power law scale}. We refer to \cite[$\S$.2.5 p.23]{Tricot} for more details. We only develop the power-law case in the following.\\

The \emph{power-law comparison scale} is defined by the family (see \cite[p.24]{Tricot}) :
\begin{equation}
\mathcal{P} =\left \{ f_{\alpha} (t)=t^{\alpha},\ \alpha >0\right \} .
\end{equation}
The exponent is obtained by looking for the quantity 
\begin{equation}
\di\frac{\ln (f_\alpha(\mu ))}{\ln (\mu )} .
\end{equation}
In order to precise the scale regime, we introduce the definition of \emph{scale range}:
\begin{definition}[Scale range]
Let $\PTT$ be a given scale sequence. A scale range between two time-scale $\TT_{m_0}$ and $\TT_{m_1}$ in $\PTT$, with $m_0$ and $m_1\in \N$ such that $m_0 <m_1$, is denoted by $[\TT_{m_0},\TT_{m_1}]$ and is defined by
\begin{equation}
[\TT_{m_0},\TT_{m_1}]=\di \bigcup_{m\geq m_0}^{m_1} \TT_m
\end{equation}
\end{definition}

Let $\PTT$ be a given scale sequence and consider a scale function $\mathbf{X}$ over $\PTT$. We then are lead to the following definition of a scale regime :
\begin{definition}[Pointwise scale regime]
Let $t\in \TT_{m_0}$ for $m_0\geq0$. The pointwise scale regime of $\mathbf{X}$ in $t$, denoted by $\alpha(\mathbf{X},t)$, is the quantity defined by
\begin{equation}
\left[\alpha(\mathbf{X},t)\right]_{\TT_m}=\di\frac{\ln \left ( \mu_m |\Delta (\mathbf{X})_m (t) |\right )}{\ln (\mu_m )},
\end{equation}
for all $m\geq m_0$.
\end{definition}

\begin{definition}[Local scale regime]
The local scale regime of $\mathbf{X}$, denoted by $\alpha(\mathbf{X})$, is defined by
\begin{equation}
\left[\alpha(\mathbf{X})\right]_{\TT_m} =\sup_{t\in \TT_m} \left[\alpha(\mathbf{X},t)\right]_{\TT_m},
\end{equation}
for all $m\geq m_0$.
\end{definition}

In applications, the important information is related to the evolution of the global scale regime when $m$ goes to infinity. Precisely, we have :

\begin{definition}[Global scale regime]
Let $m_0 ,m_1\in \N$ such that $m_0<m_1$. The global scale regime of $\mathbf{X}$ over $[\TT_{m_0},\TT_{m_1}]$ is defined by
\begin{equation}
\alpha_{\TT_{m_0},\TT_{m_1}} (\mathbf{X}) =\sup_{m_0\leq m \leq m_1} \left[\alpha(\mathbf{X})\right]_{\TT_m} .
\end{equation}
\end{definition}

\begin{remark}
These definitions are of course reminiscent of the classical pointwise and local H\"older exponent for continuous functions as defined for example in \cite{seuret}. Nevertheless, it must be noted that we do not need to assume that there exists a continuous function associated to the given scale function in order to defined such exponents. They are constructed directly on the family of discrete data which is, from our point of view, the only information that one can obtain in experimental settings.
\end{remark}

\begin{figure}[h!]
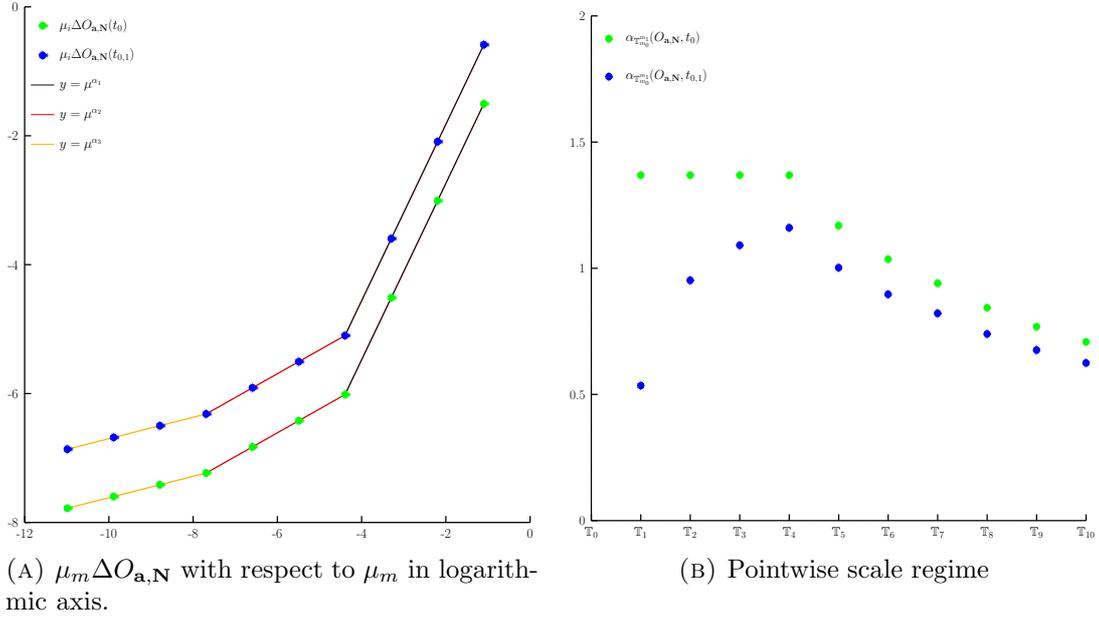

	\centering
	\begin{subfigure}[t]{0.45\textwidth}
		\centering
		\resizebox{\linewidth}{!}{\input{figure_C1.txt}}
		\caption{$\mu_m \Delta O_{\mathbf{a},\mathbf{N}}$ with respect to $\mu_m$ in logarithmic axis.}
	\end{subfigure}
	~                		
	\begin{subfigure}[t]{0.45\textwidth}
		\centering
		\resizebox{\linewidth}{!}{\input{figure_C2.txt}}
		\caption{Pointwise scale regime}
	\end{subfigure}
	\caption{MSO function with $\mathbf{a}=\{ 2/9,2/3,5/6\}$ and $\mathbf{N}=\{ 4,3,\infty\}$}
	\label{exemple1}
\end{figure}

{\bf Example :}
Consider the multiscale Okamoto's function defines by $\mathbf{a}=\{ 2/9,2/3,5/6\}$ and $\mathbf{N}=\{ 4,3,\infty\}$. Its pointwise scale regime over the scale range $[\TT_1, \TT_{10}]$ is illustrated in the Figure \ref{exemple1}. It is computed for the point $t_0$ and $t_{0,1}$ in $\TT_1$ as by construction, the pointwise scale regime of $t_{0,2}$ coincide with the one of $t_0$.
 
On this simple example, we can see the three pointwise scale power law regime over the scale range $[\TT_1, \TT_{10}]$. We also display the power law function with $\alpha_1=\frac{\ln(9/2)}{\ln(3)}$, $\alpha_2=\frac{\ln(3/2)}{\ln(3)}$ and $\alpha_3=\frac{\ln(6/5)}{\ln(3)}$ in order to compare to the slope of the pointwise scale regime. In that case, the pointwise scale regime is the same for the two points. In consequence, the global scale regime is a power law regime with $\alpha =\alpha_1$.\\

In what follows, for a general scale regime, we denote by $\mathcal{R}_{\TT_{m_0}^{m_1}}(\mathbf{X})$ the \emph{global scale regime} of $\mathbf{X}$ (or simply \emph{scale regime}) over the scale range $\left[\TT_{m_0},\TT_{m_1}\right]$.

%
%
%
%
%
%
%
%

\section{Scale dynamics}

A scale function being given we are interested in the following problems :
\begin{itemize}
	\item Can we determine if the scale function is a multiscale function ? In this case, try to determine the elementary scale actions.
	\item Assuming that we know a given scale function up to scale $m \in \N^*$, can we precise the ``best'' in some sense continuous limit model ?
\end{itemize}
To do that, we see that it is necessary to understand the behavior of the scale derivative under change of scale, i.e. \emph{scale dynamics}.\\

We begin first by studying multiscale Okamoto's function and we prove that the scale pattern complexity can be recovered by looking at the dynamics of the scale derivative. Then, we provide general transformation formula for the scale derivative between two given time-scale. 

\subsection{Scale dynamical analysis of Multiscale Okamoto's functions}
In this section, we study the behavior of the $\Delta$-derivative of multiscale Okamoto's functions under change of scale. 

\subsubsection{Scale dynamics of Okamoto's functions}

We begin with single scale Okamoto's function. Considering $\TT_0 =\{ t_0 ,t_1 \}$ and $a\in ]0,1[$. By definition of the scale Okamoto's action $O_a$ (see Definition \ref{def_scale_elem_okamoto}), we obtain the time-scale $\TT_1=\TT_{0,O_a} =\{ t_0 , t_{0,1} , t_{0,2} ,t_1 \} $ where $t_{0,i} =t_0 +i \di \frac{\mu}{3}$ for $i=1,2$ and $\mu =t_1 -t_0$. Considering the discrete function ${O_{a,\TT_0}}$ defined on $\TT_0$ and the discrete function $O_{a,\TT_1}$ defined on $\TT_1$ with the points obtained by the scale Okamoto's action. \\

We introduce the notion of {\bf reference scale regime} which we choose to be the linear scale regime, meaning that the dependence of the correction term with respect to $\mu_i$ is linear. This assumption leads to the notion of {\bf discrete reference function on $\TT_1$} associated to a given discrete function on $\TT_0$, denoted by $[{O_a}]^{\TT_1}_{\TT_0}$. 

\begin{remark}
	The choice of the linear scale regime is of course arbitrary from the mathematical point of view. However, from the physical side, it corresponds to the fact that we always compare a possible smooth one which is the standard setting in Physics. Any deviation to this framework is interpreted as the need for a new mathematical setting (non-differentiable, stochastic, etc).
\end{remark}

For Okamoto's function, the discrete reference function on $\TT_1$ is defined by $[O_a]^{\TT_1}_{\TT_0} \mid_{\TT_0 } =O_{a,\TT_0}$ and 
\begin{align*}
[O_a]^{\TT_1}_{\TT_0} (t_{0,1}) &= O_{a,\TT_0} (t_0) + \mu_1\left[\frac{O_{a,\TT_0} (t_1)-O_{a,\TT_0} (t_0)}{\mu_0}\right],\\
[O_a]^{\TT_1}_{\TT_0} (t_{0,2}) &= O_{a,\TT_0} (t_0) + 2\mu_1 \left[\frac{O_{a,\TT_0} (t_1)-O_{a,\TT_0} (t_0)}{\mu_0}\right] .
\end{align*}

By definition, we have
\begin{equation*}
\Delta O_{a,\TT_0}(t_0) = \frac{O_{a,\TT_0}(t_1)-O_{a,\TT_0}(t_0)}{t_1-t_0} \quad \text{and} \quad \Delta O_{a,\TT_1} (t_0) =\frac{3a}{\mu} \Delta O_{a,\TT_0}(t_0)
\end{equation*}
then, we obtain :

\begin{proposition}
\label{prop1_okamoto}
The $\Delta$-derivative of Okamoto's function satisfies
\begin{equation*}
\Delta O_{a,\TT_1} (t_0) = \Delta [O_a]_{\TT_0}^{\TT_1} (t_0) + \mathsf{C}(t_0)
\end{equation*}
where
\begin{equation*}
\mathsf{C}(t_0) = \left(\frac{3a}{\mu}-1\right)\Delta O_{a,\TT_0} (t_0),
\end{equation*}
\end{proposition}

In order to generalize our previous results, we introduce the following definitions :

\begin{definition}
Consider a scale function $\mathbf{F}$ on $\PTT$. We denote by $\mathbf{F}^{\star}$ the {\bf reference scale function} associated to $\mathbf{F}$  defined at scale $\TT_m$ by $F_{\TT_{m-1}}^{\TT_m}$ for all $m\geq 1$. 
\end{definition}

\begin{definition}
Let $A$ be a given scale operator on scale functions. We call {\bf scale effect induced on $A$} the difference between the action of $A$ on a given scale function and its associated reference scale function.
\end{definition}

\begin{remark}
The quantity $\mathsf{C}(t_0)$ in Proposition \ref{prop1_okamoto}, corresponds to the scale effect on $\Delta$ between $\TT_0$ and $\TT_1$.
\end{remark}

Consider now the scale sequence $\PTT$ associated with the scale or multiscale Okamoto's functions. We have :

\begin{proposition}
\label{correcokamoto}
The scale effect induced on the scale $\Delta$-derivative of Okamoto's function is given by
\begin{equation*}
\Delta O_a = \Delta O_a^\star+\mathsf{C}\left(O_a\right)
\end{equation*}
where
\begin{equation*}
\label{exprcorrokamoto}
\mathsf{C}\left(O_a\right)= \left(\frac{3a}{\mu}-1\right)\Delta O^\star_a.
\end{equation*}
Analogous formulas holds for the scale $\nabla$-derivative.
\end{proposition}
From this proposition, we recover the features of Okamoto's function given in the Theorem \ref{thm_derivability_okamoto} assuming $\mu=1$. 

\subsubsection{A scale dynamical approach to the Okamoto-Kobayashi's theorem}

The analytic properties of the Okamoto's function are related to the asymptotic behavior of the $\Delta$ and $\nabla$ derivatives and the correction term with respect to scale. Indeed, the existence of a derivative for a given point $t \in [0,1]$ can be check as follows : assume that $t\in \TT_m$ for a given $m\in \N$. The limit function admit a derivative if and only if we have 
\begin{itemize}
	\item $\di\lim_{m\rightarrow \infty}   [\Delta (O_a)]_m (t)$ and $\di\lim_{m\rightarrow \infty}   [\nabla (O_a)]_m (t)$ exist.
	\item $\di\lim_{m\rightarrow \infty} [\mathsf{C}(O_a)]_m =0$.
\end{itemize}
The first condition implies that the left and right derivatives exist at point $t$ for the limit function and the second condition implies the equality of the left and right derivative which implies derivability at point $t$.\\

An easy case is obtained when the previous conditions are satisfied from a given scale. In particular, we have :

\begin{proposition}
	The correction term $\mathsf{C}\left(O_a\right)$ vanishes if and only if $a=1/3$ or $a=1/2$.
\end{proposition}

\begin{proof}
The first point follows directly from equation \eqref{exprcorrokamoto} in Proposition \ref{correcokamoto}.

The second case follows from the following observation : in order for a given scale $\TT_m$ to have a point such that $\Delta \mathbf{F} =0$, one need to construct in the iterative procedure for a given elementary time scale as given in Definition \ref{def_scale_elem_okamoto} a configuration such that $O_a (t_{0,1} )=O_a (t_{0,2})$. This is possible if and only if $a=1/2$. 
\end{proof}

The situation between the case $a=1/3$ and $a=1/2$ is nevertheless very different. 
\begin{itemize}
\item For $a=1/3$ the correction term vanish in all point of the scale sequence. This result induces the third point of the Okamoto-Kobayashi's theorem. 

\item For $a=1/2$, the correction term vanish only on the following {\it left reduced scale sequence} obtained from the Okamoto's scale sequence taking at each step from the action on an elementary time-scale $\TT_{elem} = \{ t_0 ,t_1 \}$ the point $t_{0,1}$  from $\TT_{elem,O_a}$. We denote by $\PTT_{red}^{\leftarrow}$ this scale sequence. 
\end{itemize}

\begin{figure}[h!]
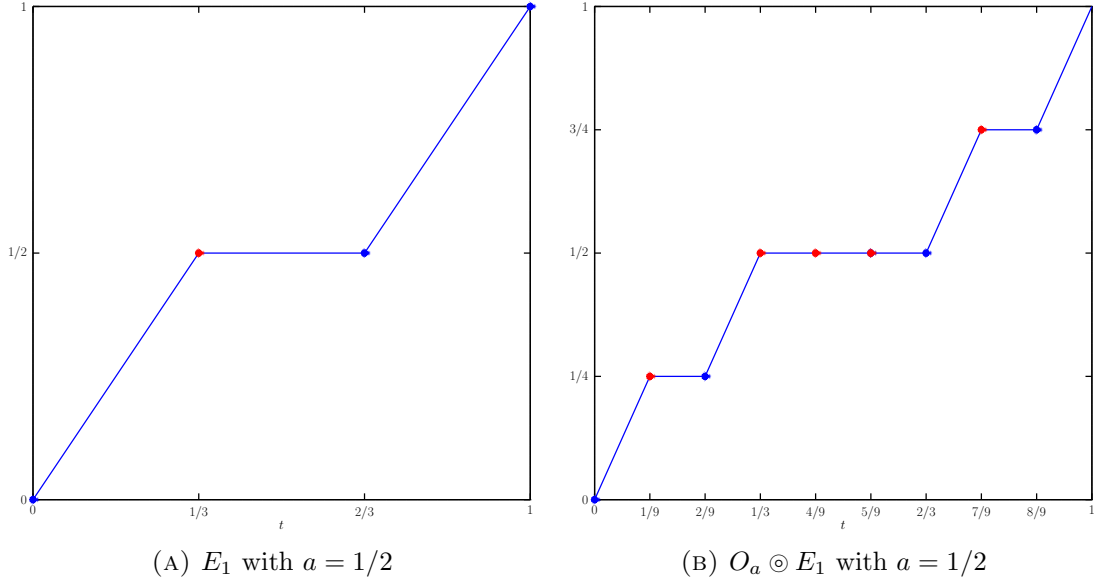

	\centering
	\begin{subfigure}[b]{0.45\textwidth}
		\centering
		\resizebox{\textwidth}{!}{\input{figure_TLR_1.txt}}
		\caption{$E_1$ with $a=1/2$}
	\end{subfigure}
	~                		
	\begin{subfigure}[b]{0.45\textwidth}
		\centering
		\resizebox{\textwidth}{!}{\input{figure_TLR_2.txt}}
		\caption{$O_a \circledcirc E_1$ with $a=1/2$}
	\end{subfigure}
	\caption{The points in red correspond to the image of the points $\PTT_{red}^{\leftarrow}$ by Okamoto's function}
\end{figure}

We have for all $t\in \PTT_{red}^{\leftarrow}$, there exists $m(t)\in \N$ such that $t\in \TT_m$ and $[\Delta (O_a )]_m (t)=0$ for all $m\geq m(t)$ .   

\begin{figure}[ht!]
	\centering
	\resizebox{0.8\textwidth}{!}{\input{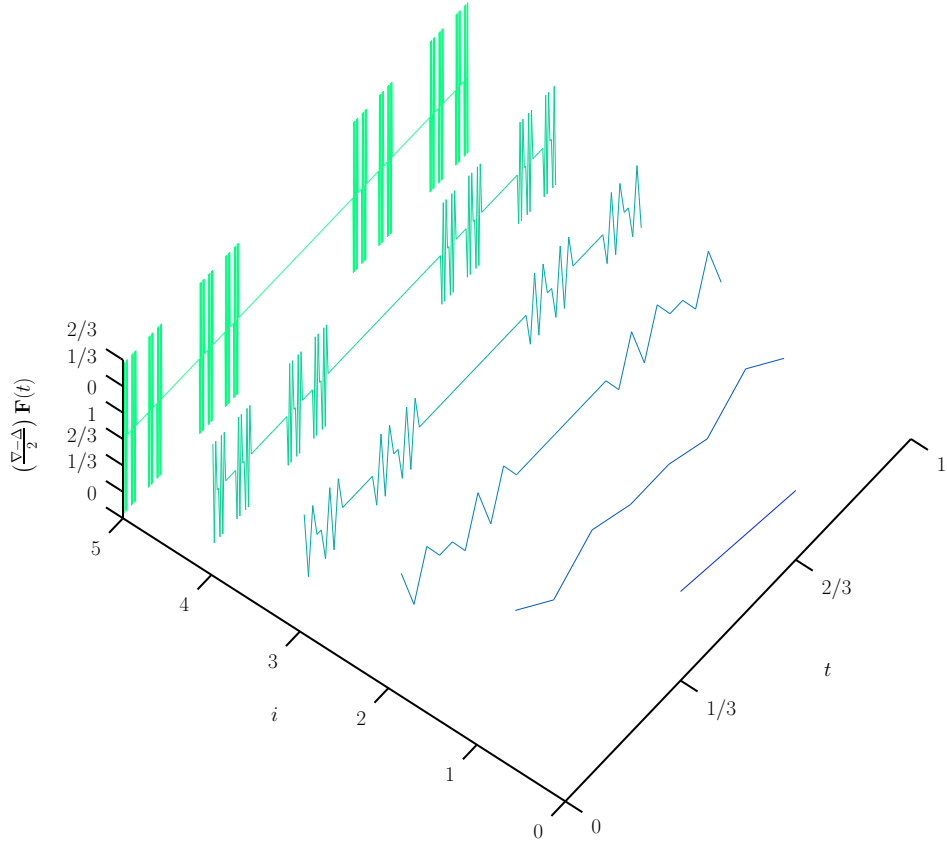}}
	\caption{Correction term for Okamoto's function $a=1/2$}
\end{figure}

This result extends easily to the $\nabla$ case changing the left reduced scale sequence for the right scale sequence obtained in the same way taking $t_{0,2}$ instead of $t_{0,1}$ in the iterative construction.\\

It must be noted that we have no information on the point of the sequence belonging to the Cantor's set $\PTT\setminus \left ( \PTT_{red}^{\leftarrow} \cup \PTT_{red}^{\rightarrow} \right )$  and defined by
\begin{equation*}
\mathbf{Cantor} = \bigcap_{p=1}^\infty \bigcap_{k=0}^{3^{p-1}-1} \left(\left[0,\frac{3k+1}{3^p}\right] \cup \left[\frac{3k+2}{3^p},1\right]\right).
\end{equation*}

In order to conclude about the derivability of the limit function, one needs to check that the $\Delta$ and $\nabla$ derivatives possess a limit when $m$ goes to infinity. We will not reproduce that computations made for example by H. Okamoto or K. Kobayashi in their papers as this is not the purpose of our work, but one can of course interpret their results in our setting.

\subsubsection{Finite scale sequence and asymptotic model identification}

Let us begin with a definition :

\begin{definition}
Let $\PTT$ be a scale sequence and $\mathbf{F}$ a scale function.  We call $F_{\infty}$ an {\it asymptotic model} for $\mathbf{F}$. 
\end{definition}

For example, $O_a$ is an asymptotic model for $\mathbf{Okamoto}$.\\

{\bf Identification problem} : {\it Assume that we have access to $\mathbf{F}$ up to the time-scale $\TT_m$. Can we  characterize the asymptotic model with these data ?}\\ 

In the Okamoto case, one needs only to identify the parameter $a$. This parameter is completely fixed using a finite set of data. As a consequence, the asymptotic model is determined knowing only a partial set of observations. Of course, this result is due to the fact that we presuppose that the scale regime will not change, i.e. that $a$ is independent of scale. This assumption can be though as an a priori point of view on the nature of the real phenomenon.\\

This result will drastically change in the multiscale case. 

\subsubsection{Scale dynamics of multiscale Okamoto's functions}

The identification problem can be more complicated than the one which is exhibited by the Okamoto's function. In the multiscale version, the problem is worse.\\

We can perform the same kind of computations for a multiscale Okamoto's function $O_{\mathbf{a}, \mathbf{N}}$ with $\mathbf{a}=(a_1 ,\dots ,a_n )$ and $\mathbf{N}=(N_1 ,\dots ,N_n)$ as in Definition \ref{def_multiscale_okamoto}. Indeed, following the same idea as previous with the scale Okamoto's functions, we obtain the following proposition :

\begin{proposition}
The scale effect induced on the scale $\Delta$-derivative of $O_{\mathbf{a}, \mathbf{N}}$ is given by
\begin{equation*}
\Delta O_{\mathbf{a}, \mathbf{N}} = \Delta O_{\mathbf{a}, \mathbf{N}}+ \mathsf{C}\left(O_{\mathbf{a}, \mathbf{N}}\right)
\end{equation*}
where 
\begin{equation*}
\mathsf{C}\left(O_{\mathbf{a}, \mathbf{N}}\right) = \left(\frac{3 s_{\mathbf{a},\mathbf{N}}}{\mu}-1\right)\Delta O^\star_{\mathbf{a}, \mathbf{N}}
\end{equation*}
and where $s_{\mathbf{a},\mathbf{N}}$ is the sequence defined in Equation \ref{sequence_multiscale}. An analogous formula holds for the scale $\nabla$-derivative.
\end{proposition}

Here again, the asymptotic properties of the model are fixed by the set $\mathbf{a}$ and $\mathbf{N}$. However, the identification problem can not be solved.

Indeed, assuming that we know the function $\mathbf{F}$ up to scale $m$, we do not know, if the sequence $s_{\mathbf{a},\mathbf{N}}$ is finite or infinite. Even in the finite case, we do not know the length of the sequence. Moreover, even if we know this length, let say $k$, and we identify already $k-1$ terms in $s_{\mathbf{a},\mathbf{N}}$ up to scale $m$, we do not know when the next change will produce.  \\

This means that, without an assumption on the asymptotic model, this is impossible to conclude from a finite scale sequence observation. This assumption is not induced by the set of accessible data and can not be checked. It comes necessarily outside of the given framework. 

\begin{remark}
In more physical terms, the choice for a particular model can then be only justified using a certain philosophical point of view on the nature of a given phenomenon which can not by definition and construction be proved by any experimental devices\footnote{This is the case for example of String theory which assumes that physical properties are coming from the description of a particular geometric object generalizing Einstein's point of view on space-time.}. 

One must noted that the previous remark is in fact valid for all kind of model which can be constructed from a given set of experimental data as, in practice, this is impossible to have access to the full scale sequence\footnote{In Physics, we have for example the well known limitation due to Heisenberg uncertainty principle.}. A limit model can not and will never represent the reality of a phenomenon as we have no possibilities to select between different admissible models up to a given scale. 
\end{remark}

\subsection{General formulas for the scale effect on $\Delta$ and $\nabla$}

\subsubsection{Passage from a scale $\TT_0$ to a scale $\TT_1$}
In this section we provide a general transformation formula for the scale derivatives between two given time-scale. \\

We want to define relations between two discrete and finite time-scales $\TT_0$ and $\TT_1$ in order to study the effect of changing scale on the scale $\Delta$-derivative and the scale $\nabla$-derivative for a given scale function. We are interested in the case where $\TT_0 \subset \TT_1$ with ${\TT_1} _{\mid \TT_0}=\TT_0$ and we consider the case where $\TT_0$ and $\TT_1$ are uniform which means their graininess function are constant. Moreover, without loss of the general idea, we consider the case where the graininess function $\mu_0$ of the time-scale $\TT_0$ and the graininess function $\mu_1$ of the time-scale $\TT_1$ are such that $\mu_0=2\mu_1$. With such considerations, we put $\TT_0=\left\{t_0,t_1\right\}$ and $\TT_1=\left\{t_0,t_\star,t_1\right\}$. \\

We consider the discrete reference function $F^{\TT_1}_{\TT_0}$ defined by
\begin{equation}
F^{\TT_1}_{\TT_0} (t_\star) = F_{\TT_0} (t_0) + \left[\frac{F_{\TT_0} (t_1)-F_{\TT_0} (t_0)}{\mu_0}\right] \mu_1 .
\end{equation}
If we compare $F_{\TT_1} \left(t_\star\right)$ with $F^{\TT_1}_{\TT_0}\left(t_\star\right)$, we can write $\Delta F_{\TT_1} (t_0)$ as
\begin{equation}
\Delta F_{\TT_1} (t_0) = \frac{F_{\TT_1}(t_\star)-F^{\TT_1}_{\TT_0} (t_\star)+F^{\TT_1}_{\TT_0} (t_\star)-F_{\TT_1}(t_0)}{\mu_1} .
\end{equation}
We obtain, using the definition of $F^{\TT_1}_{\TT_0} (t_\star)$,
\begin{equation}
\Delta F_{\TT_1} (t_0)  = \Delta F_{\TT_0}^{\TT_1} (t_0) + \frac{F_{\TT_1}(t_\star)-F^{\TT_1}_{\TT_0}(t_\star)}{\mu_1} \ .
\end{equation}

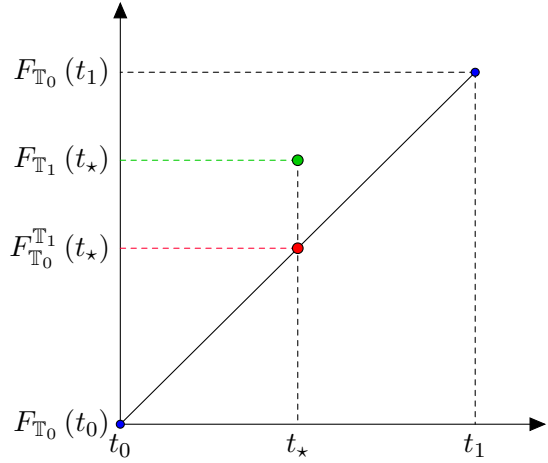
\begin{figure}[ht!]
	\centering
	\begin{tikzpicture}[line cap=round,line join=round,>=triangle 45,x=0.3\textwidth,y=0.3\textwidth]
	
	\draw[->,color=black] (0.,0.) -- (1.2,0.);
	\draw[->,color=black] (0.,0.) -- (0.,1.2);
	\clip(-0.5,-0.3) rectangle (1.2,1.2);
	
	\draw (0.,0.) node[below] {$t_0$} node[left] {\color{black} $F_{\TT_0}\left(t_{0}\right)$} --(1.,1.);
	\draw [dash pattern=on 2pt off 2pt,color=qqccqq] (0.,0.75) node[left] {\color{black} $F_{\TT_1}\left(t_\star\right)$}-- (0.5,0.75);
	\draw [dash pattern=on 2pt off 2pt,color=ffqqtt] (0.,0.5) node[left] {\color{black} $F_{\TT_0}^{\TT_1}\left(t_\star\right)$} -- (0.5,0.5);
	\draw [dash pattern=on 2pt off 2pt] (1.,1.)-- (0.,1.) node[left] {\color{black} $F_{\TT_0}\left(t_{1}\right)$};
	
	\draw [dash pattern=on 2pt off 2pt] (0.5,0.75)-- (0.5,0.) node[below] {$t_{\star}$} ;
	\draw [dash pattern=on 2pt off 2pt] (1.,1.)-- (1.,0.) node[below] {$t_{1}$} ;
	
	\draw [fill=qqqqff] (0.,0.) circle (1.5pt);
	\draw [fill=qqqqff] (1.,1.) circle (1.5pt);
	\draw [fill=ffqqqq] (0.5,0.5) circle (2.0pt);
	\draw [fill=qqccqq] (0.5,0.75) circle (2.0pt);
	\end{tikzpicture}
	\caption{Comparison between the expected point and the new point}
	\label{newpoint}  
\end{figure}

The explicit dependence of the scale behavior is shown with the last term on the right. In order to express this term as a dynamical quantity, that is to say in mean of $\Delta$ and $\nabla$, we note that by definition of the $\Delta$-derivative, $\nabla$-derivative and $F^{\TT_1}_{\TT_0} (t_\star)$, we have
\begin{equation}
\frac{F_{\TT_1} (t_\star)-F^{\TT_1}_{\TT_0} (t_\star)}{\mu_1} = \frac{\left(\nabla-\Delta\right)}{2}F_{\TT_1} (t_\star).
\end{equation}
Finally, we obtain
\begin{equation}
\Delta F_{\TT_1} (t_0)  = \Delta F_{\TT_0}^{\TT_1}  (t_0) + \frac{\left(\nabla-\Delta\right)}{2}F_{\TT_1} \circ\sigma_1(t_0).
\end{equation}
Doing the same with the $\nabla$-derivative, we obtain
\begin{equation}
\nabla F_{\TT_1} (t_1)  = \nabla F_{\TT_0}^{\TT_1} (t_1) + \frac{\left(\nabla-\Delta\right)}{2}F_{\TT_1} \circ \rho_1 (t_1).
\end{equation}
As we see, the deviation from the reference function at scale $\TT_1$ is governed by the difference between the $\Delta$ and $\nabla$ derivative of the $F_{\TT_1}$.  We give a general version of these computations in the next Section.

\subsubsection{Scale dynamics}

Consider now the scale sequence $\PTT$ where each time-scale $\TT_i\in \mathbb{T}$, $i\geq0$, has its graininess function $\mu_i$ such that $\mu_i=2\mu_{i+1}$ where $\mu_{i+1}$ is the graininess function of the time-scale $\TT_{i+1}$. \\

As we have seen in our previous computations, an important role is played by the operator $\di \frac{\nabla-\Delta}{2}$. In particular, the correction term can be expressed as an action of this operator. In order to simplify the formula in the general case, we introduce the following notation :

\begin{definition}
	[Correction term] We denote by $C_{\ltimes}$ and $C_{\rtimes}$ the left and right correction terms defined by
$C_{\ltimes} (\mathbf{F})=\di \frac{\nabla-\Delta}{2} [\ltimes (\mathbf{F} ) ]$ and $C_{\rtimes} (\mathbf{F})=\di \frac{\nabla-\Delta}{2} [\rtimes (\mathbf{F} )]$ with
\begin{equation}
\left[\ltimes ( \mathbf{F}) \right]_{\TT_m}=  
\left \{ 
\begin{array}{lll}
& \mathbf{F}\circ \rho  & \text{over}\quad \TT_{m-1}\subset \TT_m ,\\
& \mathbf{F}  & \text{over}\quad  \TT_m \setminus \TT_{m-1} ,
\end{array}
\right .
\end{equation}
and
\begin{equation}
\left[\rtimes ( \mathbf{F}) \right]_{\TT_m}=  
\left \{ 
\begin{array}{lll}
& \mathbf{F}\circ \sigma  & \text{over}\quad \TT_{m-1}\subset \TT_m ,\\
& \mathbf{F}  & \text{over}\quad  \TT_m \setminus \TT_{m-1} ,
\end{array}
\right .
\end{equation}
\end{definition}

For convenience, we introduce the following scale sign function :

\begin{definition}
We denote by $\pmb\varepsilon$ the scale sign function defined by
\begin{align}
\left[\pmb \varepsilon\right]_{\TT_m} =
\left \{ 
\begin{array}{lll}
& +1 & \text{over}\quad \TT_{m-1}\subset \TT_m ,\\
&-1 & \text{over} \quad \TT_m \setminus \TT_{m-1} .
\end{array}
\right .
\end{align}
	
\end{definition}

Considering the scale reference function $\mathbf{F}^\star$, we have the following proposition :

\begin{proposition}[Scale effect on the scale derivatives] The scale effect induced on the scale $\Delta$ and $\nabla$ derivatives of $\mathbf{F}$ is given by
\begin{align}
\Delta \mathbf{F} = \Delta \mathbf{F}^{\star} + \pmb\varepsilon \cdot \mathsf{C}_\rtimes(\mathbf{F}), \ \ \
\nabla \mathbf{F} = \Delta \mathbf{F}^{\star} + \pmb\varepsilon \cdot \mathsf{C}_\ltimes(\mathbf{F}) .
\end{align}
\end{proposition}

In many applications, we will need to have a convenient formula for the scale effect induced on the $\Delta$ and $\nabla$ derivative acting on a scale functional, i.e. a functional defined on a scale function. This is provided by the following Lemma :

\begin{lemma}
[Scale effect and chain rule]
\label{chain-rule}
Let $f :\R \times \R \rightarrow \R$ be a sufficiently smooth real valued function. We have for all scale function $\mathbf{X}$ on the scale sequence $\PTT$, the following formula
\begin{align}
\Delta f(\mathbf{T},\mathbf{X})  = \Delta f(\mathbf{T},\mathbf{X^\star}) +\pmb\varepsilon \cdot  \di\sum_{j\geq 1}  \frac{\mu^{j-1}}{j !}  [\mathsf{C}_\rtimes (\mathbf{X} ) ]^j  \di\frac{\partial^j f}{\partial x^j} (\rtimes(\mathbf{T}),\rtimes (\mathbf{X}^\star ) ).
\end{align}
and
\begin{align}
\nabla f(\mathbf{T},\mathbf{X})  = \nabla f(\mathbf{T},\mathbf{X^\star}) +\pmb\varepsilon \cdot  \di\sum_{j\geq 1}  \frac{(-1)^{j-1}\mu^{j-1}}{j !}  [\mathsf{C}_\ltimes (\mathbf{X} ) ]^j  \di\frac{\partial^j f}{\partial x^j} (\ltimes(\mathbf{T}),\ltimes (\mathbf{X}^\star ) ).
\end{align}
\end{lemma}
 
\begin{proof}
We detail the proof only for $\Delta$ as the computations are equivalent in the $\nabla$ case. We fix a given scale $\TT_m$ and we need to distinguish between points in $\TT_{m-1} \subset \TT_m$ and those in $\TT_m \setminus \TT_{m-1}$. \\

For $t\in \TT_m \setminus \TT_{m-1}$, we have $\mathbf{X} \circ \sigma (t)=\mathbf{X}^{\star} \circ \sigma (t)$ by construction, so that $f(\mathbf{T}\circ \sigma (t) ,\mathbf{X}\circ \sigma (t))=f(\mathbf{T}\circ \sigma (t) , \mathbf{X}^{\star} \circ \sigma (t))$ and $\mathbf{X} (t)=\mathbf{X}^{\star} \circ \sigma (t)- \mu \Delta \mathbf{X} (t)$. As a consequence, we obtain 
\begin{align*}
\mu \Delta f (\mathbf{T} ,\mathbf{X})(t)=& f(\mathbf{T} \circ \sigma (t),\mathbf{X}^{\star}\circ \sigma (t) )-f(\mathbf{T} (t) , \mathbf{X}(t)) ,\\
=& f(\mathbf{T} \circ \sigma (t),\mathbf{X}^{\star}\circ \sigma (t) )-f(\mathbf{T} (t) , \mathbf{X}^{\star} \sigma (t) -\mu \Delta \mathbf{X} (t)) ,\\
=& f(\mathbf{T} \circ \sigma (t),\mathbf{X}^{\star}\circ \sigma (t) )-f(\mathbf{T} (t) , \mathbf{X}^{\star} \sigma (t) -\mu \Delta \mathbf{X}^{\star} (t) -\mu \pmb\varepsilon (t) \mathsf{C}_{\rtimes}  (\mathbf{X})(t)),\\
=& f(\mathbf{T} \circ \sigma (t),\mathbf{X}^{\star} \circ \sigma(t) )-f(\mathbf{T} (t) , \mathbf{X}^{\star}  (t)  -\mu \pmb\varepsilon (t) \mathsf{C}_{\rtimes}  (\mathbf{X})(t)) .
\end{align*}
For $t\in \TT_m \setminus \TT_{m-1}$, we have $\pmb\varepsilon (t) =-1$ so that 
\begin{equation}
\mu \Delta f (\mathbf{T} ,\mathbf{X})(t) =  f(\mathbf{T} \circ \sigma (t),\mathbf{X}^{\star} (t) )-f(\mathbf{T} (t) , \mathbf{X}^{\star}  (t)  +\mu  \mathsf{C}_{\rtimes}  (\mathbf{X})(t)) .
\end{equation}
As $f$ is sufficiently smooth, we can make a Taylor expansion with respect to $(\mathbf{T} (t) , \mathbf{X^\star} (t) )$. We obtain
\begin{equation}
\mu \Delta f(\mathbf{T} ,\mathbf{X} )(t) = f(\mathbf{T} \circ \sigma (t),\mathbf{X}^\star \circ \sigma (t))-f(\mathbf{T}(t),X^{\star}(t))-\di\sum_{j\geq 1} \frac{\mu^{j}}{j !} [\mathsf{C}_{\rtimes} (\mathbf{X})(t)]^j \frac{\partial^j f}{\partial x^j} (\mathbf{T} (t) ,\mathbf{X}^\star (t) ).
\end{equation}
We then obtain for $t\in \TT_m \setminus \TT_{m-1}$ 
\begin{equation}
\Delta f(\mathbf{T},\mathbf{X})  (t)= \Delta f(\mathbf{T},\mathbf{X^\star})(t) +\pmb\varepsilon (t)\cdot  \di\sum_{j\geq 1}  \frac{\mu^{j-1}}{j !}  [\mathsf{C}_\rtimes (\mathbf{X} ) (t)]^j  \di\frac{\partial^j f}{\partial x^j} (\rtimes(\mathbf{T})(t),\rtimes (\mathbf{X}^\star )(t) ).
\end{equation}

For $t\in \TT_{m-1} \subset \TT_m$, we have $\mathbf{X}(t)=\mathbf{X}^{\star} (t)$ and $\mathbf{X}\circ \sigma (t) =\mathbf{X}^{\star}(t)+\mu \Delta \mathbf{X} (t)$. As a consequence, we obtain 
\begin{align*}
\mu \Delta f (\mathbf{T} ,\mathbf{X})(t)=& f(\mathbf{T} \circ \sigma (t),\mathbf{X} \circ \sigma (t) )-f(\mathbf{T} (t) , \mathbf{X}^{\star} (t)) ,\\
=& f(\mathbf{T} \circ \sigma (t),\mathbf{X}^{\star} (t) +\mu \Delta \mathbf{X} (t) )-f(\mathbf{T} (t) , \mathbf{X}^{\star} (t)) ,\\
=& f(\mathbf{T} \circ \sigma (t),\mathbf{X}^{\star} (t) +\mu \Delta \mathbf{X}^{\star} (t) +\mu \pmb\varepsilon (t) \mathsf{C}_{\rtimes} (t) )-f(\mathbf{T} (t) , \mathbf{X}^{\star} (t) ),\\
=& f(\mathbf{T} \circ \sigma (t),\mathbf{X}^{\star} \circ \sigma (t) +\mu \pmb\varepsilon (t) \mathsf{C}_{\rtimes} (t) )-f(\mathbf{T} (t) , \mathbf{X}^{\star}  (t)) .
\end{align*}
For $t\in \TT_{m-1} \subset \TT_{m}$, we have $\pmb\varepsilon (t) =1$ so that 
\begin{equation}
\mu \Delta f (\mathbf{T} ,\mathbf{X})(t)=
f(\mathbf{T} \circ \sigma (t),\mathbf{X}^{\star} \circ \sigma (t) +\mu \mathsf{C}_{\rtimes} (t) )-f(\mathbf{T} (t) , \mathbf{X}^{\star}  (t)) .
\end{equation}
As $f$ is sufficiently smooth, we can make a Taylor expansion with respect to $(\mathbf{T} \circ \sigma (t) , \mathbf{X^\star} \circ  \sigma (t) )$. We obtain
\begin{equation}
\mu \Delta f(\mathbf{T} ,\mathbf{X} )(t) = f(\mathbf{T} \circ \sigma (t),\mathbf{X}^\star \circ \sigma (t))-f(\mathbf{T}(t),X^{\star}(t))+\di\sum_{j\geq 1} \frac{\mu^{j}}{j !} [\mathsf{C}_{\rtimes} (\mathbf{X})(t)]^j \frac{\partial^j f}{\partial x^j} (\mathbf{T} \circ \sigma (t) ,\mathbf{X}^\star \circ \sigma (t) ).
\end{equation}	
We then obtain for $t\in \TT_{m-1} \subset \TT_{m}$ 
\begin{equation}
\Delta f(\mathbf{T},\mathbf{X})  (t)= \Delta f(\mathbf{T},\mathbf{X^\star})(t) +\pmb\varepsilon (t)\cdot  \di\sum_{j\geq 1}  \frac{\mu^{j-1}}{j !}  [\mathsf{C}_\rtimes (\mathbf{X} ) (t)]^j  \di\frac{\partial^j f}{\partial x^j} (\rtimes(\mathbf{T})(t),\rtimes (\mathbf{X}^\star )(t) ).
\end{equation}
\end{proof}

\section{Asymptotic differential operator}

The previous scale analysis can be used to define a new class of differential operator on the asymptotic class of functional set covered by limit of scale functions. These operators generalize the Box derivative introduced in \cite{cresson_greff} et the It\^o calculus for stochastic processes (see \cite{oksendal}). The basic idea is to used the natural decomposition of a given scale function having a fixed scale regime in its ``regular'' part and its deviation part which is responsible for the scale regime.
	
\subsection{Extension and decomposition of scale functions}

Let $m_0<m_1$ and  $[\TT_{m_0} ,\TT_{m_1}]$ be a scale range over which a scale function $\mathbf{X}$ possesses a scale regime of order $0<\alpha <1$.  Then, we have :
\begin{itemize}
	\item We call {\bf extension} of $\mathbf{X}$ and we denote by $\mathbf{X}_{ext}$ a scale function such that $\mathbf{X} \mid_{[\TT_{m_0},\TT_{m_1}]} =\mathbf{X}$ and $\mathbf{X}_{ext}$ possesses a scale regime of order $\alpha$ over $[\TT_{m_1 +1} ,\TT_{\infty}]$. Then $\mathbf{X}_{ext ,\infty}$ belongs to $H^{\alpha} ([a,b],\R )$.
	
	\item Moreover, we have $\mu^{j-1}_m \left(\frac{(\nabla -\Delta)}{2}\mathbf{X}\right)^j$ of order $\mu_m^{j\alpha-1}$ for all $m>m_0$ and $j\geq 1$. 
	
	\item By construction, we have $[\mathbf{X}^{\star}]_{ext}$ which possesses a linear scale regime and is associated to a piecewise $C^1$ function denoted by $X_{\infty}^{\star}$.
\end{itemize}
As a consequence, we can write $\mathbf{X}_{ext} =\mathbf{X}_{ext}^{\star} + \mathbf{D}_{ext}$ where $\mathbf{D}_{ext} = \mathbf{X}_{ext} -\mathbf{X}_{ext}^{\star}$. The asymptotic limit of $\mathbf{X}_{ext}^{\star}$ denoted $X_{\infty}$ is decomposed as 
\begin{equation}
X_{\infty} =X^{\star}_{\infty} + D_{\infty} ,
\end{equation}
where $X_{\infty}^{\star}$ is a piecewise differentiable function and the deviation part from this differentiable behavior is given by $D_{\infty}$ which is a particular way of decomposing $X_{\infty}$ into a regular and non regular part  

\begin{remark}
In \cite{cresson_greff} this operation is made directly on a very special functional space for which the non regular part is fixed in a given class. This allows the authors in this case to defined a projection which gives well defined regular and non regular part. 
\end{remark} 

\subsection{Asymptotic differential operator}

The $\Delta$ and $\nabla$ derivatives are asymptotically equivalent to the classical derivative as long as the scale regime is linear. However, they do not possess asymptotic limit on scale function possessing a scale regime of order $0<\alpha <1$. Using the previous extension/decomposition of a scale function over a given scale regime, we give a meaning to a differential operator acting on $X_{\infty}$ by taking advantage from the fact that a part of the decomposition admits a classical derivative.

\begin{definition}
Let $\mathbf{X}$ be a given scale function and $\mathcal{R}_{\TT_{m_0} ,\TT_{m_1}}$ a scale regime with $m_0<m_1$. We denote by $\mathbf{X}_{ext}$,  $\mathbf{X}^{\star}_{ext}$ and $\mathbf{D}_{ext}$ the extension and decomposition of $\mathbf{X}$ associated to $\mathcal{R}_{\TT_{m_0} ,\TT_{m_1}}$ and $X_{\infty}$, $X_{\infty}^{\star}$, $D_{\infty}$ the asymptotic limit and decomposition of $\mathbf{X}$. We then define the operator $\Delta_{\infty}$ acting on $X_{\infty}$ as 
\begin{equation}
\Delta_{\infty} [X_{\infty} ]=\di\frac{d}{dt} X_{\infty}^{\star} .
\end{equation}
\end{definition}

As a consequence, the operator $\Delta_{\infty}$ extracts the derivative of the regular part of $X_{\infty}$. \\

Another way to formalize this operation is to introduce the operator $\textsf{reg}$ on scale function such that $\mbox{reg} (\mathbf{X}_{ext}) =X_{ext}^{\star}$. In this case, we have 
\begin{equation}
\Delta_{\infty} [X_{\infty}] = \Delta_{\infty} [\textsf{reg} (\mathbf{X}_{ext})] = \di\frac{d}{dt} X_{\infty}^{\star}.
\end{equation}

\begin{remark}
This definition must be compared with the definition of the Box derivative in \cite{cresson_greff} and also the definition of the Nelson's forward and backward derivatives over stochastic processes in \cite{nelson2001}.
\end{remark}

For applications, we need as usual the behavior of this operator over composition of functions. In order to have explicit formula we introduce a special functional space over scale functions : 

\begin{definition}
Let $\mathbf{X}$ be a given scale function and $\mathcal{R}_{\TT_{m_0} ,\TT_{m_1}}^{\alpha}$ a power law scale regime of order $0<\alpha <1$ with $m_0<m_1$. We denote by $\mathbf{X}_{ext}$,  $\mathbf{X}^{\star}_{ext}$ and $\mathbf{D}_{ext}$ the extension and decomposition of $\mathbf{X}$ associated to $\mathcal{R}_{\TT_{m_0} ,\TT_{m_1}}^{\alpha}$. We denote by $C_{\Delta ,\lambda_+}$ (resp. $C_{\nabla ,\lambda_-}$) the set of scale functions such that for $j_{\alpha} =E(1/\alpha)$ there exists $\lambda_+$ (resp. $\lambda_+$) such that
\begin{equation}
\label{consistence}
\lim_{m\rightarrow  \infty} \mu_m^{j_{\alpha}-1} [\mathsf{C}_{\rtimes} (\mathbf{T},\mathbf{X}_{ext})]^{j_{\alpha}} =\lambda^{j_{\alpha}}_+ \quad \left(resp. \ \lim_{m\rightarrow  \infty} \mu_m^{j_{\alpha}-1} [\mathsf{C}_{\ltimes} (\mathbf{T},\mathbf{X}_{ext})]^{j_{\alpha}} =\lambda^{j_{\alpha}}_-\right).
\end{equation}
\end{definition}

We have the following result : 

\begin{proposition}
\label{prop_ito}
Let $\mathbf{X}$ be a given scale function and $\mathcal{R}_{\TT_{m_0} ,\TT_{m_1}}^{\alpha}$ a power law scale regime of order $0<\alpha <1$ with $m_0<m_1$. We denote by $\mathbf{X}_{ext}$,  $\mathbf{X}^{\star}_{ext}$ and $\mathbf{D}_{ext}$ the extension and decomposition of $\mathbf{X}$ associated to $\mathcal{R}_{\TT_{m_0} ,\TT_{m_1}}$ and $X_{\infty}$, $X_{\infty}^{\star}$, $D_{\infty}$ the asymptotic limit and decomposition of $\mathbf{X}$. Assume that $\mathbf{X} \in C_{\Delta ,\lambda_+}$. We then define the operator $\Delta_{\infty}$ acting on $f(t,X_{\infty})$ as 
\begin{equation}
\Delta_{\infty}  f(t,X_{\infty}) : = \di\frac{d^+}{dt} f(t,X_{\infty}^{\star} ) +  \frac{\lambda^{j_{\alpha}}_+}{j_{\alpha} !}   \di\frac{\partial^{j_{\alpha} } f}{\partial x^{j_{\alpha}}} (t, X_{\infty}^{\star} ) .
\end{equation}
\end{proposition}

\begin{proof}
We have 
\begin{equation}
\Delta f(\mathbf{T},\mathbf{X}_{ext})  = \Delta f(\mathbf{T},\mathbf{X}_{ext}^\star ) +\pmb\varepsilon \cdot  \di\sum_{j\geq 1}  \frac{\mu^{j-1}}{j !}  [\mathsf{C}_\rtimes (\mathbf{X}_{ext} ) ]^j  \di\frac{\partial^j f}{\partial x^j} (\rtimes(\mathbf{T}),\rtimes (\mathbf{X}_{ext}^\star ) ).
\end{equation}
As the quantity $\frac{\mu^{j-1}}{j !}  [\mathsf{C}_\rtimes (\mathbf{X}_{ext} ) ]^j  \di\frac{\partial^j f}{\partial x^j} (\rtimes(\mathbf{T}),\rtimes (\mathbf{X}_{ext}^\star ) )$ admits a linear scale regime for $j\geq j_{\alpha}$ by assumption, we have that 
\begin{equation}
\mbox{reg} \left [ \di\sum_{j\geq 1}  \frac{\mu^{j-1}}{j !}  [\mathsf{C}_\rtimes (\mathbf{X}_{ext} ) ]^j  \di\frac{\partial^j f}{\partial x^j} (\rtimes(\mathbf{T}),\rtimes (\mathbf{X}_{ext}^\star ) ) \right ] = 
\di\sum_{j\geq j_{\alpha} }  \frac{\mu^{j-1}}{j !}  [\mathsf{C}_\rtimes (\mathbf{X}_{ext} ) ]^j  \di\frac{\partial^j f}{\partial x^j} (\rtimes(\mathbf{T}),\rtimes (\mathbf{X}_{ext}^\star ) ) .
\end{equation}
However, the asymptotic of these quantities are trivial for $j>j_{\alpha}$ so that the limit reduces to 
\begin{equation}
 \frac{1}{j_{\alpha} !} \lambda^{j_{\alpha}}_+  \di\frac{\partial^{j_{\alpha} } f}{\partial x^{j_{\alpha}}} (t, X_{\infty}^{\star} ) .
\end{equation}
\end{proof}
In the same way we have
\begin{proposition}
\label{prop_ito_bis}
Let $\mathbf{X}$ be a given scale function and $\mathcal{R}_{\TT_{m_0} ,\TT_{m_1}}^{\alpha}$ a power law scale regime of order $0<\alpha <1$ with $m_0<m_1$. We denote by $\mathbf{X}_{ext}$,  $\mathbf{X}^{\star}_{ext}$ and $\mathbf{D}_{ext}$ the extension and decomposition of $\mathbf{X}$ associated to $\mathcal{R}_{\TT_{m_0} ,\TT_{m_1}}$ and $X_{\infty}$, $X_{\infty}^{\star}$, $D_{\infty}$ the asymptotic limit and decomposition of $\mathbf{X}$. Assume that $\mathbf{X} \in C_{\nabla ,\lambda_-}$. We then define the operator $\nabla_{\infty}$ acting on $f(t,X_{\infty})$ as 
\begin{equation}
\nabla_{\infty}  f(t,X_{\infty}) : = \di\frac{d^-}{dt} f(t,X_{\infty}^{\star} ) -  \frac{\lambda^{j_{\alpha}}_- }{j_{\alpha} !}  \di\frac{\partial^{j_{\alpha} } f}{\partial x^{j_{\alpha}}} (t, X_{\infty}^{\star} ) .
\end{equation}
\end{proposition}

\subsection{Extension to complex scale functions and comparison with the Box derivative}

Following the strategy exposed in \cite{cresson2005} and further developed in \cite{cresson_greff}, we can extend the asymptotic scale operator to complex scale functions assuming linearity of the operator :

\begin{definition}[Asymptotic Box derivative]
The asymptotic Box derivative, denoted by $\Box_{\infty}$ is the linear operator defined over complex valued scale functions by
\begin{equation}
\Box_{\infty} = \di\frac{1}{2} \left (\Delta_{\infty} +\nabla_{\infty} \right ) + i\frac{\eta }{2} \left ( 
\Delta_{\infty} -\nabla_{\infty} \right ) ,  
\end{equation}
where $i^2 =-1$ and $\eta =\{ -1,1,-i,i \}$.
\end{definition}

In \cite{cresson_greff}, the definition is very similar and is in some sense equivalent even if not formulated in the same formalism. The main problem in \cite{cresson_greff} was to extract from the one parameter family of averaging of a given continuous functions, some information which can be encoded in a kind of derivative. The one parameter family associated to a given function can be clearly seen as a kind of scale function. The scale regime is then responsible for the divergence of the left and right derivative which is taking into account in \cite{cresson_greff} using some projection operator on the set of convergent function depending on the parameter. This is the role of the regularization operator in our definition.\\

Nevertheless, we point out that the main point in our approach is that no asymptotic object need to exist ! Te asymptotic object is constructed using the extension procedure and does not necessarily corresponds to the real asymptotic (if any) of the scale function.\\

Using Proposition \ref{prop_ito} and \ref{prop_ito_bis}, we have :
\begin{proposition}
\label{prop_ito_box}
Let $\mathbf{X}$ be a given scale function and $\mathcal{R}_{\TT_{m_0} ,\TT_{m_1}}$ a scale regime with $m_0<m_1$. We denote by $\mathbf{X}_{ext}$,  $\mathbf{X}^{\star}_{ext}$ and $\mathbf{D}_{ext}$ the extension and decomposition of $\mathbf{X}$ associated to $\mathcal{R}_{\TT_{m_0} ,\TT_{m_1}}$ and $X_{\infty}$, $X_{\infty}^{\star}$, $D_{\infty}$ the asymptotic limit and decomposition of $\mathbf{X}$. Assume that $\mathbf{X} \in C_{\Delta ,\lambda_+}\cap C_{\nabla ,\lambda_-}$. We then define the operator $\Box_{\infty}$ acting on $f(t,X_{\infty})$ as 
\begin{equation}
\Box_\infty  f(t,X_{\infty}) := \frac{\Box}{\Box t}  f(t,X_{\infty})+   \frac{\lambda_\alpha}{j_{\alpha} !} \di\frac{\partial^{j_{\alpha} } f}{\partial x^{j_{\alpha}}} (t, X_{\infty}^{\star} ) .
\end{equation}
where
\begin{equation}
\lambda_\alpha = \left[\left(\lambda^{j_{\alpha}}_+ - \lambda^{j_{\alpha}}_-\right) + i\eta\left(\lambda^{j_{\alpha}}_++\lambda^{j_{\alpha}}_-\right)\right]
\end{equation}
\end{proposition}

One can compare this result with the corresponding result in \cite{cresson_greff}.

\section{Asymptotic models of scale equations}
\label{continuous}
The previous formalism can be used to derived what is the adapted {\bf asymptotic model} associated to a given {\bf scale differential equation}. 

\begin{definition}
We call scale differential equation any formal relation of the form 
\begin{equation}
F(\mathbf{T} ,\mathbf{X} ,\Delta ,\nabla ) =0 .
\end{equation}
\end{definition} 

As an example, a particular class of second order scale differential equation is given by 
\begin{equation}
\label{scale_delta_2ord}
\nabla \circ \Delta \mathbf{X} = f(\mathbf{T} , \mathbf{X}) .
\end{equation}

These models are usually constructed from a given {\bf macroscopic model} corresponding for us to an initial scale $\TT_0$. This macroscopic model is in many cases associated to a classical differential or partial differential equation like for example the classical Newton's equation in classical mechanics. This representation must be thought as a particular asymptotic model assuming a very special scale regime, namely the linear one. \\

We give two examples which will be used in the Section concerning applications of this idea.

\subsection{Linear scale regime}

Take a scale sequence $\PTT$ and the set of scale functions having a linear scale regime. In that case, using the scale dynamics equation, we observe that 
\begin{equation}
\di\lim_{m\rightarrow \infty} \mathsf{C}_{\rtimes} (\mathbf{X}) =\di\lim_{m\rightarrow \infty}  \mathsf{C}_{\ltimes} (\mathbf{X}) =0 ,
\end{equation}
When $m$ goes to infinity, we have by construction that $\TT_m$ goes to a closed interval $[a,b]$ and any discrete function belonging to the functional set $C(\TT_m ,\R )$ is converging to $C ([a,b],\R )$. The operator $\Delta$ and $\nabla$ are converging to the classical right and left derivatives.\\

For the previous scale equation, we must assume that $\mathbf{X} \in C^2_{\nabla,\Delta} (\PTT)$, i.e. $\mathbf{X}$ is scale $\Delta$ differentiable over $\PTT$ and $\Delta \mathbf{X}$ is scale $\nabla$ differentiable over $\PTT$. If $\mathbf{X}$ satisfies moreover that 
$\lim_{m\rightarrow \infty} \Delta (\mathbf{X})_{\TT_m}$ and  $\lim_{m\rightarrow \infty} \nabla (\mathbf{X})_{\TT_m}$ exist, then the condition of linear scale regime implies that $\mathbf{X}$ is converging to $X_{\infty} \in C^2 ([a,b],\R )$.\\
 
As a consequence, under the linear scale regime assumption, the asymptotic model associated to the previous second order scale differential equation is simply the classical second order differential equation
\begin{equation}
\di\frac{d^2 x}{dt^2} = f(t,x) .
\end{equation} 
Of course, this assumption is known to be false. The scale regime, even in the classical setting, changes drastically when going to the microscopic scale. The question is then to understand what is the correct asymptotic model when the scale regime is changing. Our formalism provide such an answer, taking always the deviation with respect to the linear scale regime as reference. In other words, this means that we compare the behavior in different scale regime with respect to the macroscopic one took as a reference. 

\subsection{Fractional scale regime} 

If we replace the linear regime by a fractional one of order $0<\alpha <1$ the situation changes drastically.\\

If the initial scale equation is equivalent under the change of variable $f$ to the equation
\begin{equation}
\Delta f(\mathbf{T},\mathbf{X}_{ext}) =G(\mathbf{T},\mathbf{X}_{ext}^{\star}) .
\end{equation}
Then its asymptotic continuous model is given by 
\begin{equation}
\Delta_{\infty}  f(t,X_{\infty}) =G(t,X_{\infty}^{\star} ),
\end{equation}
which can be written as a classical (partial) differential equation 
\begin{equation}
\label{modasymp}
\di\frac{d^+}{dt} f(t,X_{\infty}^{\star} ) +  \frac{1}{j_{\alpha} !} \lambda^{j_{\alpha}}_+  \di\frac{\partial^{j_{\alpha}} f}{\partial x^{j_{\alpha}}} (t, X_{\infty}^{\star} ) =G(t,X_{\infty}^{\star} ).
\end{equation}

We can resume the previous approach by the following diagram : 

\begin{figure}[ht!]
	\centering
	\resizebox{0.5\textwidth}{!}{
		\begin{tikzpicture}
		
		\node(E0) at (0,6) {\centering $\Delta f(\mathbf{T},\mathbf{X})=\Delta f(\mathbf{T},\mathbf{X}^\star)+\mathsf{C}_\mathcal{R}(\mathbf{T},\mathbf{X})$};
		
		\node(E1) at (0,3) {$\Delta f(\mathbf{T},\mathbf{X}_{ext})=\Delta f(\mathbf{T},\mathbf{X}^\star_{ext})+\mathsf{C}_\mathcal{R}(\mathbf{T},\mathbf{X}_{ext})$};
		
		\node(E2) at (0,0) {$\Delta_\infty f(t,X^\star_\infty)=\frac{d^+}{dt}f(t,X^\star_\infty)+\mathsf{C}_{\infty,\mathcal{R}}(t,X^\star_\infty)$};
		
		\draw[->] (E0) to node[align=center,fill=white,rectangle,draw]{\textbf{Extension} \\ $\mathbf{X}_{ext}$} (E1);
		\draw[->] (E1) to node[align=center,fill=white,rectangle,draw]{\textbf{Asymptotic model}\\$m\rightarrow \infty$} (E2);
		
		\end{tikzpicture}
	}
	\caption{Diagram showing the approach of the extension and the passage to an asymptotic model}
\end{figure}
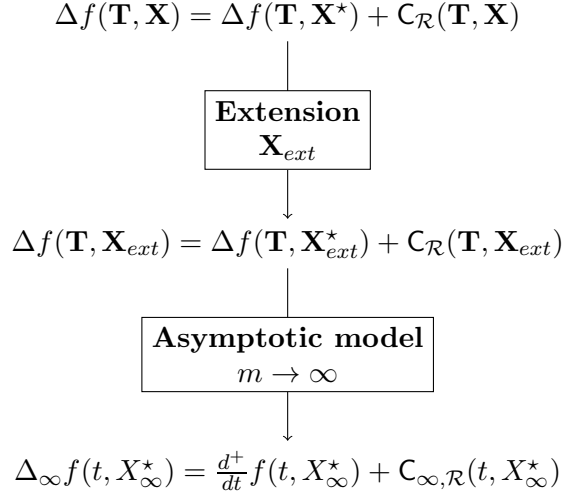

The important point is that equation (\ref{modasymp}) is a classical (partial) differential equation. As a consequence, this construction allows us to interpret perturbations of classical models as effects due to different scale regime over a single scale invariant equation. An example of this situation is given in the next Section with the Newton's equation.

\subsection{Asymptotic versus scale equations}

The natural object coming from modeling in Physics, Biology or any modeling based on experimental data can be formalized in the framework of scale equations, dealing only with a finite number of quantities. However, most of the present models are formulated in the context of the differential calculus or different generalization like the stochastic calculus of Schwartz distributions. As we have seen, these continuous models corresponds in our language to special asymptotic model of a given scale system. The asymptotic character of these models is in general completely lost in the current formulation of Physics, etc, which can be called the {\bf fundamental continuous modeling assumption} which is, by no way, explicit in the existing literature.\\

{\bf Fundamental continuous modeling assumption} : {\it Assume that a given system exhibit a given scale regime. A continuous model is obtained assuming that the observed scale regime is in fact valid up to infinity.}\\

This fundamental assumption is very useful when one is trying to interpret some very strange result coming from the analysis or numerical analysis of some models. This is in particular clear in the context of fluid mechanics where the governing equation are precisely written under an asymptotic assumption which relate the macroscopic and microscopic behavior. \\

Assume, to see the difficulties induced by this assumption, that the scale regime change at very small scale. Then, doing numerical simulation of the corresponding continuous model, one must wait for the emergence of nonphysical or strange solutions. The same is true for the analysis of current continuous models, when one studies solutions exhibiting some scale structures. The Navier-Stokes equation or Euler equation are in this respect very illustrative. Simulations of the Navier-Stokes equation induce most of the time unwanted solution with no physical meaning. The problem is then to choose between the validity of the equation itself and the validity of the numerical method. Some paradoxical solutions are known for the Euler equation which have no physical meaning but can be explicitly constructed. For an overview of results and questions in fluid mechanics, we refer to \cite{gerard,villani}. The construction makes use of a recursive and scale behavior. The question is then to choose between again the equation or the definition of generalized solutions that one must introduce to cover for example turbulent behavior. \\

Again, all these problems are due to the fact that an asymptotic model has been used base on the fundamental continuous modeling assumption which is not explicit in the presentation of these models. A given continuous model is valid as long as the fundamental continuous modeling assumption can be in some extent justified. This is not the case for fluid mechanics for example, at least for turbulent regime where scale structures are assumed to appear at all scales. 

\section{Applications to partial differential equations : the Diffusion equation and the Schr\"odinger equation}
\label{applications}

If different scale regime are present, we obtain with the previous formalism different asymptotic models with different correction terms view as deviation from the linear scale. This is illustrated as follows:

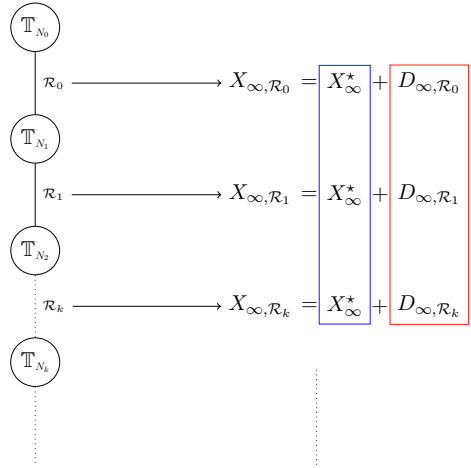
\begin{figure}
	\centering
	\tikzset{
		mystyle/.style={
			circle,
			inner sep=0pt,
			text width=8mm,
			align=center,
			draw=black,
			fill=white
		}
	}
	\resizebox{0.4\textwidth}{!}{
		\begin{tikzpicture}
		
		\node[mystyle](E0) at (0,0) {$\mathbb{T}_{\scalebox{0.5}{$N_0$}}$};
		\node[mystyle](E1) at (0,-2) {$\mathbb{T}_{\scalebox{0.5}{$N_1$}}$};
		\node[mystyle](E2) at (0,-4) {$\mathbb{T}_{\scalebox{0.5}{$N_2$}}$};
		\node[mystyle](E3) at (0,-6) {$\mathbb{T}_{\scalebox{0.5}{$N_k$}}$};
		\node(E4) at (0,-8) {};
		
		\draw[-] (E0) to node[align=center,right](b1) {\tiny$\mathcal{R}_0$} (E1);
		\draw[-] (E1) to node[align=center,right](b2) {\tiny$\mathcal{R}_1$} (E2);
		\draw[dotted] (E2) to node[align=center,right](b3) {\tiny$\mathcal{R}_{k}$} (E3);
		\draw[dotted] (E3) to node[align=center,right] {} (E4);
		
		\node(E00) at (4,|- b1) {$X_{\infty,\mathcal{R}_0}$};
		\node(E11) at (4,|- b2) {$X_{\infty,\mathcal{R}_1}$};
		\node(E22) at (4,|- b3) {$X_{\infty,\mathcal{R}_k}$};
		
		\node(E000) at (5.5,|- b1) {$X^\star_\infty$};
		\node(E111) at (5.5,|- b2) {$X^\star_\infty$};
		\node(E222) at (5.5,|- b3) {$X^\star_\infty$};

		\node(E0000) at (7,|- b1) {$D_{\infty,\mathcal{R}_0}$};
		\node(E1111) at (7,|- b2) {$D_{\infty,\mathcal{R}_1}$};
		\node(E2222) at (7,|- b3) {$D_{\infty,\mathcal{R}_k}$};
		
		\draw[->] (b1) to (E00);
		\draw[->] (b2) to (E11);
		\draw[->] (b3) to (E22);
		
		\draw[white] (E00) to node[align=center,black] {=} (E000);
		\draw[white] (E11) to node[align=center,black] {=} (E111);
		\draw[white] (E22) to node[align=center,black] {=} (E222);
		
		\draw[white] (E000) to node[align=center,black] {+} (E0000);
		\draw[white] (E111) to node[align=center,black] {+} (E1111);
		\draw[white] (E222) to node[align=center,black] {+} (E2222);
		
		\draw[blue!100!black!90] (E000.west |- ,|- E000.north) rectangle (E222.east |-,|- E222.south);
		\draw[red!100!black!90] (E0000.west |- ,|- E0000.north) rectangle (E2222.east |-,|- E2222.south);
		
		\node(F1) at (5,-6) {};
		\node(F2) at (5,-8) {};
		\draw[dotted] (F1) to (F2);
		\end{tikzpicture}
	}
	\caption{Illustrations of different continuous models with different scale regime. The part in blue corresponds to the classical or regular part. The part in red corresponds to all the deviations to the regular part depending on the scale regime $\mathcal{R}_k$ chosen.}
\end{figure}

Such a result is particularly interesting when studying a {\bf scale invariant equation}. Indeed, a single scale invariant equation will provide a possibly {\bf infinite number of asymptotic continuous models} corresponding to different scale regimes. The fact that different continuous models exist for a given physical phenomenon is then understood as the fact that the phenomenon exhibit different scale regimes during scaling. As a consequence, the {\bf multiplicity of continuous models} does not imply that no {\bf universal equation} underlies the phenomenon {\bf in the scale framework}. Each continuous equation is valid in its own domain of scales. 

\subsection{Scale Newton's equation and the diffusion equation}

We consider the classical equation obtained by Newton to describe the dynamical behavior of a particle of mass $m$ under the action of a force deriving from a potential $U$. Precisely, we call {\it Newton's equation} the following ordinary differential equation of order $2$ 
\begin{equation}
\di\frac{d^2 x}{dt^2} = U'(x) .
\end{equation}
The {\bf scale Newton's equation} is defined by
\begin{equation}
\label{scale_newton_delta}
\nabla \circ \Delta \mathbf{X} = U' (\mathbf{X}) .
\end{equation}
The choice of this equation as a scale analogue of the classical Newton's equation is supported by the following results : For each scale, this equation corresponds to the {\it variational embedding} in the sense of \cite{cresson2012} of the classical Newton's equation which means that the solution of the scale Newton's equation coincide with the extremal of the scale embedding of the classical Lagrangian structure under the time-scale calculus of variations. \\
\begin{remark}
The class of scale equations \eqref{scale_delta_2ord} corresponds to the choice of $\Delta$ as the derivative of the position. As the scale Newton's equation \eqref{scale_newton_delta} is defined in this class, we have also its $\nabla$ version given by
\begin{equation}
\label{scale_newton_nabla}
\Delta \circ \nabla \mathbf{X} = U' (\mathbf{X}).
\end{equation}
\end{remark}
Let us assume that there exists a function $\psi (t,x)$ such that 
\begin{equation}
\Delta \mathbf{X} =-2\gamma \di\frac{\partial \ln ( \psi )}{\partial x} (\mathbf{T},\mathbf{X}),
\end{equation}
and 
\begin{equation}
\Delta \mathbf{X}=-\nabla \mathbf{X} .
\end{equation}
We then have the following theorem :

\begin{theorem}
Assume that $\mathbf{X}$ possess a scale regime of order $1/2$. Then the asymptotic continuous model of the scale Newton's equation under the change of variables $\psi$ is given by 
\begin{equation}
\di {\frac{\partial \psi }{\partial t}}  + \left(\gamma +\frac{\lambda^2_-}{2}\right) \di\frac{1}{\psi} \left ( \di\frac{\partial \psi}{\partial x} \right )^2   -\di {\frac{ \lambda^2_-}{2}}  \di {\partial^2 \psi\over \partial x^2} +\frac{1}{2\gamma}U \psi=0 .
\end{equation}
\end{theorem}
Specializing to the subspace defined by 
\begin{equation}
\gamma =-\frac{\lambda_-^2}{2} ,
\end{equation}
we deduce an interpretation of the diffusion equation as the dynamical behavior of the classical Newton equation on a particular scale regime :
\begin{theorem}[Diffusion equation versus Newton equation]
Assume that $\mathbf{X}$ possess a scale regime of order $1/2$. Then the asymptotic continuous model of the scale Newton's equation under the change of variables $\psi$ is given by 
\begin{equation}
\di {\frac{\partial \psi }{\partial t}} = \di {\frac{ \lambda^2_-}{2}}  \di {\partial^2 \psi\over \partial x^2} +\frac{1}{\lambda_-^2}U \psi.
\end{equation}
\end{theorem}

\begin{proof}

By Proposition \ref{prop_ito}, we have

\begin{align*}
\label{cal1}
\nabla_\infty \di \left (\di {\frac{\partial \ln (\psi )}{\partial x}} (t,X_\infty(t)) \right ) = &\di {\frac{\partial }{\partial t}} \left (\di {\partial \ln (\psi )\over \partial x} \right ) (t,X^\star_\infty(t)) \\
& + \di {\frac{d^- X^\star_\infty(t)}{dt}} \di \frac{\partial}{\partial x} \left (\di {\partial \ln (\psi )\over \partial x}\right ) (t,X^\star_\infty(t)) \\
& -\di {1\over 2} \lambda^2_- \di {\partial^2 \over \partial x^2} \left (\di {\partial \ln (\psi )\over \partial x}\right ) (t,X^\star_\infty(t)).
\end{align*}
As  $\Delta \mathbf{X}=-\nabla \mathbf{X}$, we obtain 
\begin{equation*}
\di {\frac{d^- X^\star_\infty(t)}{dt}} = -\di {\frac{d^+ X^\star_\infty(t)}{dt}} =2\gamma \di\frac{\partial \ln (\psi )}{\partial x} (t,X^{\star}_{\infty} (t)) .
\end{equation*}
Replacing in the equation, we deduce 
\begin{align*}
\nabla\infty \di \left (\di {\frac{\partial \ln (\psi )}{\partial x}} (t,X_\infty(t)) \right ) = &\di {\frac{\partial }{\partial t}} \left (\di {\partial \ln (\psi )\over \partial x} \right ) (t,X^\star_\infty(t)) \\
& + 2\gamma \di\frac{\partial \ln (\psi )}{\partial x} (t,X^{\star}_{\infty} (t))
 \di \frac{\partial}{\partial x} \left (\di {\partial \ln (\psi )\over \partial x}\right ) (t,X^\star_\infty(t)) \\
& -\di {1\over 2} \lambda^2_- \di {\partial^2 \over \partial x^2} \left (\di {\partial \ln (\psi )\over \partial x}\right ) (t,X^\star_\infty(t)).\\
= &\di {\frac{\partial }{\partial t}} \left (\di {\partial \ln (\psi )\over \partial x} \right ) (t,X^\star_\infty(t)) + \gamma \di\frac{\partial}{\partial x} \left [ \di\frac{1}{\psi^2} \left ( \di\frac{\partial \psi}{\partial x} \right )^2 \right ] (t,X^{\star}_{\infty} (t))\\
& -\di {1\over 2} \lambda^2_- \di {\partial^2 \over \partial x^2} \left (\di {\partial \ln (\psi )\over \partial x}\right ) (t,X^\star_\infty(t)).\\
= &\di\frac{\partial}{\partial x}
\left [ 
\di {\frac{\partial }{\partial t}} \ln (\psi )  (t,X^\star_\infty(t)) + \gamma  \di\frac{1}{\psi^2} \left ( \di\frac{\partial \psi}{\partial x} \right )^2  (t,X^{\star}_{\infty} (t)) \right .\\
& \left . -\di {1\over 2} \lambda^2_- \di {\partial \over \partial x} \left (\di {\partial \ln (\psi )\over \partial x}\right ) (t,X^\star_\infty(t)) \right ] .\\
\end{align*}
As a consequence, developing the derivatives of $\ln \psi$, we have
\begin{align*}
\nabla_\infty \di \left (\di {\frac{\partial \ln (\psi )}{\partial x}} (t,X_\infty(t)) \right )= &\di\frac{\partial}{\partial x}
\left [ \di\frac{1}{\psi} 
\di {\frac{\partial \psi }{\partial t}}  + \gamma  \di\frac{1}{\psi^2} \left ( \di\frac{\partial \psi}{\partial x} \right )^2   -\di {1\over 2} \lambda^2_- \left ( \di\frac{1}{\psi} \di {\partial^2 \psi\over \partial x^2} - \di\frac{1}{\psi^2 } \di \left ( {\partial \psi \over \partial x}\right )^2 \right ) \right ](t,X^\star_\infty(t)) .\\
= &\di\frac{\partial}{\partial x}
\left [ \di\frac{1}{\psi} 
\di {\frac{\partial \psi }{\partial t}}  + \left(\gamma +\frac{\lambda^2_-}{2}\right) \di\frac{1}{\psi^2} \left ( \di\frac{\partial \psi}{\partial x} \right )^2   -\di {\frac{ \lambda^2_-}{2}} \di\frac{1}{\psi} \di {\partial^2 \psi\over \partial x^2} \right ](t,X^\star_\infty(t)) .\\
\end{align*}
Using the scale Euler-Lagrange equation, we deduce that 
\begin{equation*}
\di\frac{\partial}{\partial x}
\left [ \di\frac{1}{\psi} 
\di {\frac{\partial \psi }{\partial t}}  + \left(\gamma +\frac{\lambda^2_-}{2}\right) \di\frac{1}{\psi^2} \left ( \di\frac{\partial \psi}{\partial x} \right )^2   -\di {\frac{ \lambda^2_-}{2}} \di\frac{1}{\psi} \di {\partial^2 \psi\over \partial x^2}+ \frac{1}{2\gamma}U \right ](t,X^\star_\infty(t))  =0 .
\end{equation*}
As a consequence, we obtain the following partial differential equation
\begin{equation}
\di {\frac{\partial \psi }{\partial t}}  + \left(\gamma +\frac{\lambda^2_-}{2}\right) \di\frac{1}{\psi} \left ( \di\frac{\partial \psi}{\partial x} \right )^2   -\di {\frac{ \lambda^2_-}{2}}  \di {\partial^2 \psi\over \partial x^2} +\frac{1}{2\gamma}U \psi=0 .
\end{equation}
This concludes the proof.
\end{proof}

\subsection{Box scale Newton's equation and the Schr\"odinger equation}

In \cite{cresson_greff}, the Box derivative allows to recover the nonlinear and linear \emph{Schr\"odinger equation}. This is also the case here using the definition of the Box derivative in the context of scale dynamics. As in the previous section for the diffusion equation, let us assume that
\begin{equation}
\label{hypo}
\Delta \mathbf{X}=-\Delta \mathbf{X} .
\end{equation}
and there exists a function $\psi (t,x)$ such that 
\begin{equation}
\Delta \mathbf{X} =-2\gamma \di\frac{\partial \ln ( \psi )}{\partial x} (\mathbf{T},\mathbf{X}).
\end{equation}
Using the definition of the Box derivative, it leads to
\begin{equation}
\di {\frac{\Box X^\star_\infty(t)}{\Box t}} = -2i\gamma \di\frac{\partial \ln (\psi )}{\partial x} (t,X^{\star}_{\infty} (t)) .
\end{equation}
Combining the $\Delta$ and $\nabla$ version of the {\bf scale Newton's equation} with the assumption given in the Equation \eqref{hypo}, we obtain what we called the {\bf Box scale Newton's equation} as 
\begin{equation}
\frac{\Box}{\Box t} \left( \frac{\Box \mathbf{X}}{\Box t}\right)   = U'(\mathbf{X}).
\end{equation}

\begin{theorem}
Assume that $\mathbf{X}$ possess a scale regime of order $1/2$. Then the asymptotic continuous model of the Box scale Newton's equation under the change of variables $\psi$ is given by 
\begin{equation}
-2i\gamma\left(\di {\frac{\partial \psi }{\partial t}}  - \left(i\gamma +\frac{\lambda_2}{2}\right) \di\frac{1}{\psi} \left ( \di\frac{\partial \psi}{\partial x} \right )^2  +\di {\frac{ \lambda_2}{2}}  \di {\partial^2 \psi\over \partial x^2}\right) +U \psi=0 .
\end{equation}
\end{theorem}
Specializing to the subspace defined by 
\begin{equation}
\gamma =\frac{\hbar}{2}, \ \lambda_-^2=\lambda_+^2 = \hbar^2
\end{equation}
and with $\eta =-1$, we deduce an interpretation of the Schr\"odinger equation as the dynamical behavior of the classical Newton equation on a particular scale regime :
\begin{theorem}[Schr\"odinger equation versus Newton equation]
Assume that $\mathbf{X}$ possess a scale regime of order $1/2$. Then the asymptotic continuous model of the scale Newton's equation under the change of variables $\psi$ is given by 
\begin{equation}
i\hbar \di {\frac{\partial \psi }{\partial t}} + \di {\frac{ \hbar^2}{2}}  \di {\partial^2 \psi\over \partial x^2} =U \psi.
\end{equation}
\end{theorem}

\begin{proof}
The proof follows from the same kind of computations as for the diffusion equation only replacing $\Delta_{\infty}$ by the asymptotic Box derivative.
\end{proof}

\section{Conclusion}
From the modeling problem, we develop in this paper a formalism called \emph{scale dynamics}, which allows us to deal with multiple finite and discrete time-scale. Precisely, we give explicit formulas showing how behave the scale derivatives under change of scale. This is illustrated with the extension of Okamoto's function, called scale and multiscale Okamoto's function. We also discuss why there exists different continuous models associated with the same scale equation and we apply it on the Newton's equation. The formalism developed in this paper goes beyond the time-scale calculus which is naturally included as a special case.

\section*{Notations}

\noindent\begin{tabular}{@{} p{.32\textwidth}@{} p{.68\textwidth}@{}}
	
	$\mathbb{T}$ & Finite and discrete time scale\\
	
	$\mathbb{T}_{elem}$ & Elementary time scale i.e. time-scale with two elements\\
	
	$Elem(\mathbb{T})$ & Decomposition of $\mathbb{T}$ in elementary time scale\\
	
	$\PTT$ & Scale sequence\\
	
	$C(\mathbb{T},\R)$ & Space of discrete functions\\
	
	$C(\PTT,\R)$ & Space of scale functions\\
	
	$\mathbf{F}$ & A scale function \\
	
	$A \circledcirc F$ & Action of an operator $A$ on a discrete function $F$\\
	
	$\mathbb{T}_{elem,A}$ & Time scale obtained with the action of $A$ \\
	
	$O_a$ & Okamoto's function \\
	
	$O_{\mathbf{a},\mathbf{N}}$ & Multiscale Okamoto's function \\
	
	$O_{\mathbf{a},\mathbf{N},m}$ & Multiscale Okamoto's function of order $m$\\
	
	$\textbf{Okamoto}$ & Set of multiscale Okamoto's functions \\
	
	$[\TT_{m_0},\TT_{m_1}]$ & Scale range from $\TT_{m_0}$ to $\TT_{m_1}$ with $m_0<m_1$ \\
	
	$\alpha(X,t)$ & Pointwise power-law scale regime in $t$ \\
	
	$\alpha(X)$ & Local power-law scale regime\\
	
	$\alpha_{\TT_{m_0},\TT_{m_1}}(X)$ & Global power-law scale regime\\
	
	$\mathcal{R}_{\TT_{m_0},\TT_{m_1}}(X)$ & Global scale regime\\
	
	$\mathbf{X}_{ext}$ & Extension of the scale function $\mathbf{X}$ \\
	
	$\mathbf{X}^\star_{ext}$ & Reference scale function of $\mathbf{X}_{ext}$ \\
	
	$X_\infty$ & Asymptotic model of the scale function $\mathbf{X}$ \\
	
	$X^\star_\infty$ & Regular part of the asymptotic model of the scale function $\mathbf{X}$ \\
	
	$\sigma$ & Forward jump\\
	
	$\rho$ & Backward jump \\
	
	$\mu$ & Forward graininess function \\
	
	$\nu$ & Backward graininess function \\
	
	$\Delta$ & Forward derivative \\
	
	$\Delta_\infty$ & Asymptotic forward derivative \\
	
	$\nabla$ & Backward derivative \\
	
	$\nabla_\infty$ & Asymptotic forward derivative \\
	
	$\displaystyle \int\Delta t$ & Cauchy $\Delta$-integral \\
	
	$\frac{\Box}{\Box t}$ & Box derivative \\

	$\Box_\infty$ & Asymptotic Box derivative \\
	
\end{tabular} 

\bibliographystyle{plain}

\end{document}